\documentclass[twoside,twocolumn,9pt]{article}
\usepackage{extsizes}
\usepackage[super,sort&compress,comma]{natbib} 
\usepackage[version=3]{mhchem}
\usepackage[left=1.5cm, right=1.5cm, top=1.785cm, bottom=2.0cm]{geometry}
\usepackage{balance}
\usepackage{mathptmx}
\usepackage{sectsty}
\usepackage{graphicx} 
\usepackage{lastpage}
\usepackage[format=plain,justification=justified,singlelinecheck=false,font={stretch=1.125,small,sf},labelfont=bf,labelsep=space]{caption}
\usepackage{float}
\usepackage{fancyhdr}
\usepackage{fnpos}
\usepackage[english]{babel}
\addto{\captionsenglish}{%
  
}
\usepackage{array}
\usepackage{droidsans}
\usepackage{charter}
\usepackage[T1]{fontenc}
\usepackage[usenames,dvipsnames]{xcolor}
\usepackage{setspace}
\usepackage[compact]{titlesec}
\usepackage{hyperref}

\usepackage{epstopdf}

\definecolor{cream}{RGB}{222,217,201}

\usepackage[normalem]{ulem}

\begin{document}

\pagestyle{fancy}
\thispagestyle{plain}
\fancypagestyle{plain}{
\renewcommand{\headrulewidth}{0pt}
}

\makeFNbottom
\makeatletter
\renewcommand\LARGE{\@setfontsize\LARGE{15pt}{17}}
\renewcommand\Large{\@setfontsize\Large{12pt}{14}}
\renewcommand\large{\@setfontsize\large{10pt}{12}}
\renewcommand\footnotesize{\@setfontsize\footnotesize{7pt}{10}}
\makeatother

\renewcommand{\thefootnote}{\fnsymbol{footnote}}
\renewcommand\footnoterule{\vspace*{1pt}%
\color{cream}\hrule width 3.5in height 0.4pt \color{black}\vspace*{5pt}} 
\setcounter{secnumdepth}{5}

\makeatletter 
\renewcommand\@biblabel[1]{#1}            
\renewcommand\@makefntext[1]%
{\noindent\makebox[0pt][r]{\@thefnmark\,}#1}
\makeatother 
\renewcommand{\figurename}{\small{Fig.}~}
\sectionfont{\sffamily\Large}
\subsectionfont{\normalsize}
\subsubsectionfont{\bf}
\setstretch{1.125} 
\setlength{\skip\footins}{0.8cm}
\setlength{\footnotesep}{0.25cm}
\setlength{\jot}{10pt}
\titlespacing*{\section}{0pt}{4pt}{4pt}
\titlespacing*{\subsection}{0pt}{15pt}{1pt}

\fancyfoot{}
\fancyfoot[LO,RE]{\vspace{-7.1pt}\includegraphics[height=9pt]{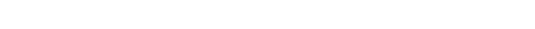}}
\fancyfoot[CO]{\vspace{-7.1pt}\hspace{13.2cm}\includegraphics{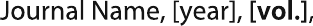}}
\fancyfoot[CE]{\vspace{-7.2pt}\hspace{-14.2cm}\includegraphics{head_foot/RF}}
\fancyfoot[RO]{\footnotesize{\sffamily{1--\pageref{LastPage} ~\textbar  \hspace{2pt}\thepage}}}
\fancyfoot[LE]{\footnotesize{\sffamily{\thepage~\textbar\hspace{3.45cm} 1--\pageref{LastPage}}}}
\fancyhead{}
\renewcommand{\headrulewidth}{0pt} 
\renewcommand{\footrulewidth}{0pt}
\setlength{\arrayrulewidth}{1pt}
\setlength{\columnsep}{6.5mm}
\setlength\bibsep{1pt}

\makeatletter 
\newlength{\figrulesep} 
\setlength{\figrulesep}{0.5\textfloatsep} 

\newcommand{\topfigrule}{\vspace*{-1pt}%
\noindent{\color{cream}\rule[-\figrulesep]{\columnwidth}{1.5pt}} }

\newcommand{\botfigrule}{\vspace*{-2pt}%
\noindent{\color{cream}\rule[\figrulesep]{\columnwidth}{1.5pt}} }

\newcommand{\dblfigrule}{\vspace*{-1pt}%
\noindent{\color{cream}\rule[-\figrulesep]{\textwidth}{1.5pt}} }

\makeatother

\twocolumn[
  \begin{@twocolumnfalse}
{\includegraphics[height=30pt]{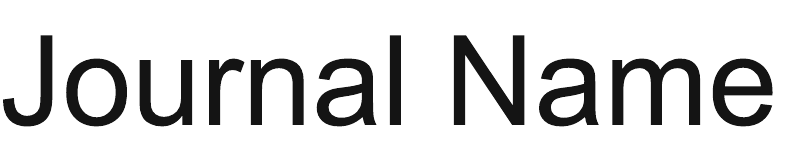}\hfill\raisebox{0pt}[0pt][0pt]{\includegraphics[height=55pt]{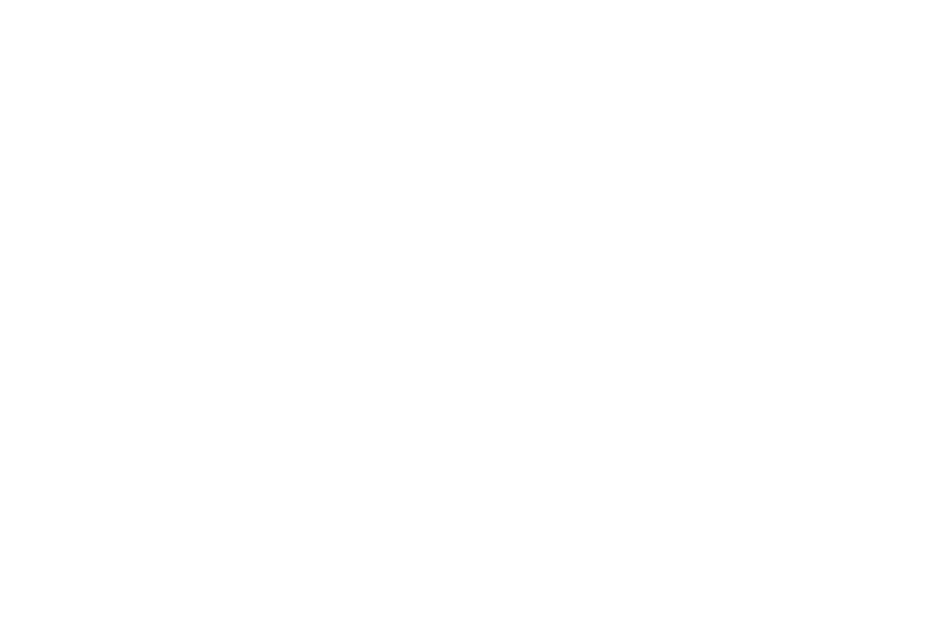}}\\[1ex]
\includegraphics[width=18.5cm]{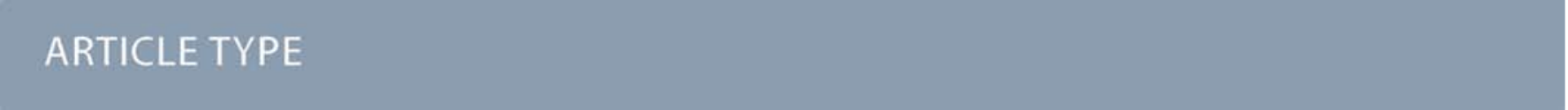}}\par
\vspace{1em}
\sffamily
\begin{tabular}{m{4.5cm} p{13.5cm} }

\includegraphics{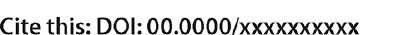} & \noindent\LARGE{\textbf{Single Active Ring Model$^\dag$}} \\
\vspace{0.3cm} & \vspace{0.3cm} \\

 & \noindent\large{Emanuel F. Teixeira,$^{\ast}$\textit{$^{a}$} Heitor C. M. Fernandes,\textit{$^{a\ddag}$} and Leonardo G. Brunnet\textit{$^{a}$}} \\

\includegraphics{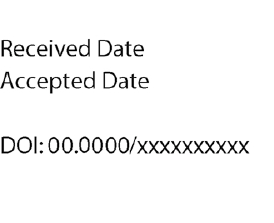} & \noindent\normalsize{Cellular tissue behavior is a multiscale problem. At the cell level, out of equilibrium, biochemical reactions drive physical cell-cell interactions in a typical active matter process. Cell modeling computer simulations are a robust tool to explore the countless possibilities and test hypotheses.
Here, we introduce a two dimensional, extended active matter model for biological cells. A ring of interconnected self-propelled particles represents the cell. Translational modes, rotational modes, and mixtures of these appear as collective states. Using analytic results derived from active Brownian particles, we identify effective characteristic time scales for ballistic and diffusive movements. Finite-size scale investigation shows that the ring diffusion increases linearly with its size when in collective movement. A study on the ring shape reveals that all collective states are present even when bending forces are weak. In that case, when in translational mode, the collective velocity aligns with the largest ring's direction in a spontaneous polarization emergence.} \\

\end{tabular}

 \end{@twocolumnfalse} \vspace{0.6cm}

  ]

\renewcommand*\rmdefault{bch}\normalfont\upshape
\rmfamily
\section*{}
\vspace{-1cm}


\footnotetext{\textit{$^{a}$~Instituto de Física, Universidade Federal do Rio Grande do Sul, CP 15051, CEP 91501-970 Porto Alegre - RS, Brazil; E-mail: teixeiraemanuel9@gmail.com, heitor.fernandes@ufrgs.br, leon@if.ufrgs.br}}

\footnotetext{\dag~Electronic Supplementary Information (ESI) available: [details of any supplementary information available should be included here]. See DOI: 10.1039/cXsm00000x/}



\section{Introduction}


 Active matter systems are constructed based on interacting elements that move using energy or mass fluxes, resulting in an emerging complex behavior~\cite{rama2010}. Cells in living tissues are physically active elements playing the role prescribed by the underlying biochemical system.
Wound healing, morphogenesis, and tumor evolution are essential processes in living organisms and motivate research on phenomena related to multi-cellular organization~\cite{LEVIN2012243,COCHETESCARTIN20172827}.
Computational modeling may identify essential physical ingredients responsible for tissue regenerative behavior~\cite{marcheti2013,vanliedekerke2015}.  Hypothesis concerning cell segregation, such as Differential Adhesion~\cite{Steinberg401} and Different Velocities~\cite{Jones1989}, were simulated based on simple point-like, active matter models~\cite{belmonte2008,beatrici2011}. However, more sophisticated hypotheses taking into account cell cortex tension, such as  Superficial Contraction~\cite{harris1976}, can not be explored using these models.

Cell movement depends on an internal actin fiber structure, which polymerizes or depolymerizes as different membrane parts reach substrate regions with fluctuating rigidity or experiment chemical gradients\cite{Aber1980,Schwarz2013}. 
To describe the physical forces under such a fine structure, extended cell models come into play. Monte-Carlo based model, such as GGH\cite{graner1992}, or more recently, the vertex model\cite{bi2016,barton2017} and the phase-field approach\cite{shao2012,Blan2013}, came in to fill this gap. These models use energy fluctuations or field equations in their description. We present here a complementary approach based on active molecular dynamics.

In this work, we present an extended cell model for active systems able to contemplate several features of other models while keeping its simplicity and physical appeal. A model cell is constructed based on a set of active particles connected by springs and subject to a bending potential, forming a ring. Here we show the different dynamical states a single ring may assume using well-known order parameters to identify collective translation~\cite{vicsek}. We also calculate a slightly modified version of the group angular momentum~\cite{erdmann2005noise} as an order parameter to characterize the single ring collective rotation. We study the ring diffusion and frame it in the context of active Brownian particles using known analytical solution limits\cite{cates2013} and experimental observations\cite{howse2007}. Finally, we investigate its shape and size change under different parameters using the gyration tensor.

The paper is structured as follows: In Sec.~\ref{sec.model}, we present the model and simulation details; in Sec.~\ref{sec.results} results for quantities used to characterize the behavior collective motion, Sec.~\ref{sec.motion}, mean square displacement and its effective parameters, Sec.~\ref{sec.msd}, and ring's morphology, Sec.~\ref{sec.morfo}; In Section~\ref{sec.conc}, we present our conclusions and summarize the results. 

\section{Model} \ 
\label{sec.model}

\begin{figure}[!h]
\centering
\includegraphics[width=0.9\columnwidth]{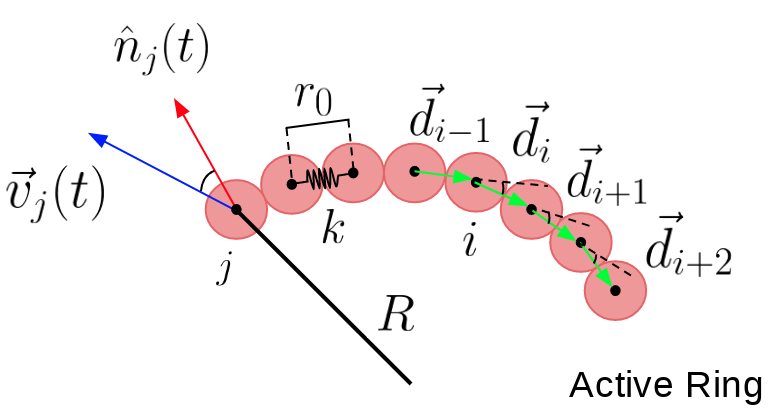}
\caption{
Sketch of a ring segment  illustrating vectors, angles and distances used in the model definition. See text for details.
}
\label{Ring scheme}
\end{figure}

We model the cell as 
a ring formed by $N$ active particles held together by $N$ bonds and subject to bending forces (see Fig.~\ref{Ring scheme}). This last interaction plays two roles: prevent ring collapse and determine its shape, in the absence of other forces, while allowing membrane fluctuations.  Our two-dimensional system lies in a square box with periodic boundary conditions. We neglect inertial effects supposing a  low-Reynolds-number regime~\cite{purcell1977life,bechinger2016active}. The overdamped equations~\cite{szabo} governing each particle dynamics are
\begin{eqnarray}
\label{1}
  \frac{\mathrm{d}}{\mathrm{d} t} \vec{r}_{i}(t)  &=& v_{0}\,\hat{n}_{i} -\mu \sum_{i\sim j }^{}\nabla U(\vec{r}_{ij}) + \sqrt{2\,D_{T}} \,\vec{\chi}_{i}(t)  \\
\label{2}
\frac{\mathrm{d}}{\mathrm{d} t} \theta_{i}(t) &=& \frac{1}{\tau'}  \arcsin{\left(\hat{n}_{i}(t)\times \frac{\vec{v}_{i}(t)}{|\vec{v_{i}}(t)|}\cdot \hat{e}_z\right)}+ \sqrt{2\,D_{R}}\,\xi_{i}(t)
\end{eqnarray}
where $\vec{r}_{i}(t) = (x_{i}(t) , y_{i}(t))$ denotes the $i$-th particle's position at time $t$, $\mu$ its mobility, and $v_{0}$  the magnitude of the self-propelling velocity along with its orientation, $\hat{n}_{i}(t) = (\cos \theta_{i}(t) , \sin \theta_{i}(t) )$. The direction of the self-propelling velocity described by the angle $\theta_{i}(t)$, relaxes towards  $\vec{v}_{i}\equiv d\vec {r}_i/dt$ within a characteristic time $\tau'$, while also experiencing angular Gaussian white noise  $\xi_{i}$ with zero-mean and second moment $\left \langle \xi_{i}(t_{1})\xi_{j}(t_{2}) \right \rangle = \delta_{ij}\delta (t_{1}-t_{2})$ independently for each particle at each time-step. $D_{R}$  is the rotational diffusion coefficient and defines a typical timescale,  $\tau_R\equiv 1/D_R$, for changes due to angular noise.
When translational noise is present, each self-propelled particle position is subject to Gaussian noise with zero-mean and variance $\left \langle \vec{\chi}_{i}(t_{1}).\vec{\chi}_{j}(t_{2}) \right \rangle = 2\delta_{ij}\delta (t_{1}-t_{2})$. $D_{T}$ is the translational diffusion coefficient and defines a characteristic timescale,  $\tau_T\equiv\sigma^2/D_T$, for a particle to diffuse a length of the order of its size, $\sigma$.

The derivatives of the  inter-particle bond, bending, and excluded-volume (EV) potentials generate the forces on each particle,
\begin{equation}
\label{3}
    U = U_{bond} + U_{bend} + U_{EV}.
\end{equation}
For the bond term we use a harmonic potential,
\begin{equation}
\label{4}
    U_{bond} = \frac{k}{2}\sum_{j=i}^{i+1} (\,|\vec{d}_{j}| - r_{0}\,)^{2},
\end{equation}
where $\vec{d}_{j} = \vec{r}_{j} - \vec{r}_{j-1}$ is the bond vector connecting consecutive particles in the ring (see Fig.~\ref{Ring scheme}), $k$ is the spring constant and $r_{0}$ is the equilibrium bond length. We introduce a bending potential to control the ring rigidity~\cite{allen2017},
\begin{equation}
\label{5}
  U_{bend} = \frac{k_{b}}{2}\sum_{j=i}^{i+2} \frac{\vec{d}_{j}.\vec{d}_{j-1}}{|\vec{d}_{j}||\vec{d}_{j-1}|},
\end{equation}
 $k_{b}$ is the bending rigidity. We model the excluded-volume interaction among particles using a Weeks–Chandler–Anderson potential (WCA),
\begin{equation}
\displaystyle
\label{wca}
    U_{EV} = \frac{\epsilon}{12}\begin{cases}
 \left ( \sigma/r_{ij} \right )^{12} - \left ( \sigma/r_{ij} \right )^{6},& \text{ if } r_{ij}<2^{1/6}\sigma \\ 
0, & \text{ if } r_{ij}\geq 2^{1/6}\sigma
\end{cases}
\end{equation}
$r_{ij} = |\vec{r}_{i}(t) - \vec{r}_{j}(t)|$, $\epsilon$, $\sigma$ are the distance between particles $i$ and $j$, characteristic exclusion volume energy and effective diameter of a given particle, respectively.

\subsection{Control Parameters}
It is useful to identify dimensionless parameters to control system dynamics. From  Eq.~\ref{1} and Eq.~\ref{2}, we define
\begin{eqnarray}
\label{BT}
B \equiv  \frac{\tau_{R}}{\tau_{T}},
\end{eqnarray}
when $\tau_T\rightarrow\infty$, translational noise is irrelevant and $B \rightarrow 0$. Parameter $B$ measures the relative importance between angular and translational noise.
Another dimensionless parameter is the rotational Péclet number~\cite{martin2018collective}, 
\begin{equation}
\label{7}
   Pe \equiv  \frac{\tau_R}{2\tau_0},
\end{equation}
with 
$\tau_0\equiv\sigma/v_0$. $Pe$ relates rotational diffusion and movement's persistence time. 
 
Also, we follow the work by Duman and collaborators\cite{duman2018collective}, and define the flexure number as
\begin{equation}
\label{8}
    Fn \equiv  \frac{N\sigma^2 (\tau_0^{-1} +\tau_{T}^{-1}) }{\mu \, k_{b}},
\end{equation}
parameters  $\tau_0^{-1}$ and $\tau_T^{-1}$ play the role of deforming forces, while the bending force, $k_{b}$, tends to restore the circular shape. 

\begin{figure}[!h]
\centering
\includegraphics[width=0.7\columnwidth]{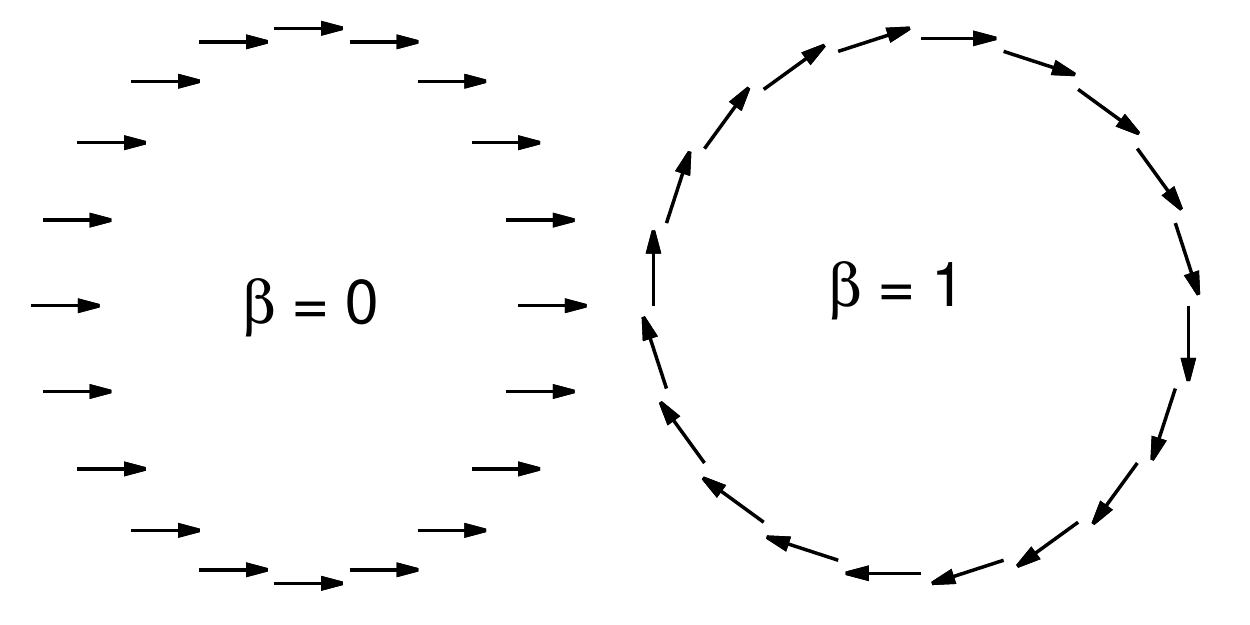} 
\caption{
Initialization of positions and velocities used in simulation. Particles are disposed in a circle with radius $R$. Self-propelled velocities direction, $\hat{n}_{i}$, initialization is determined through parameter $\beta$; $\beta=0$ sets translational motion configuration and $\beta=1$ sets a rotational one. 
}
\label{initialization}
\end{figure}

Initial conditions specification follows equations,
\begin{eqnarray}
\label{eq.ic}
    \vec{r}_{i}(t=0) &=& R \, \cos \phi_{i} \, \hat{e}_x + R \, \sin \phi_{i} \, \hat{e}_y,\\
    \hat{n}_{i}(t=0) &=& \frac{\left[ (1-\beta) + \beta \, \sin \phi_{i} \right] \, \hat{e}_x -\beta \, \cos \phi_{i} \, \hat{e}_y}{\sqrt{\left[ (1-\beta) +
    \beta \, \sin \phi_{i} \right]^{2} + \beta^{2} \, \cos^{2} \phi_{i}}} ,
\end{eqnarray}
where $R = Nr_{0}/2\pi$, $r_{0} = 2^{1/6}\sigma$, $\phi_{i} = (i-1)r_{0}/R$ and $i = 1, 2, ... N$. We initialize the ring with a circular shape.
Parameter $\beta$ defines the initial polarization, $\hat{n}_{i}$, for each particle, $\beta=1$ implies a circular polarization and $\beta=0$ a parallel one, see Fig.~\ref{initialization}.

We integrate the basic dynamic equations,  Eq.~\ref{1} and Eq.~\ref{2}, using Euler method with a time step in the range $\Delta t/\tau_{0} = [10^{-4}:2.10^{-3}]$. Along this work, we use $k = 10\epsilon/\sigma^{2}$, $v_{0} = 0.1$, $\epsilon = 1$, $\mu = 1$ and $\sigma = 1$. We also define $\tau\equiv\tau'/\tau_0$. This choice of parameters guarantees  $k$ sufficiently large to render bond length close $r_{0}$. Through the rest of paper, time is in units of $\tau_0$ and we refer to reduced time, $t/\tau_0$, just as $t$ for sake of simplification.

\subsection{Order Parameters}

A well defined velocity correlation is the signature for collective motion~\cite{czirok1996formation,strombom2011collective}. Here we identify both translational and rotational orders.   
To quantify translation we use~\cite{vicsek,szabo}
\begin{equation}
\label{9}
  \varphi(t) =  \frac{1}{N} \left | \sum_{i}^{N}\frac{\vec{v}_{i}(t)}{|\vec{v}_{i}(t)|}  \right |,
\end{equation}
which measures whether self-propelled velocities are aligned promoting translational collective movement.
To quantify rotation we use
\begin{equation}
\label{10}
        \Gamma(t) \equiv \frac{1}{N} \left | \sum_{i}^{N} \frac{{\vec{r}}_{i,CM}(t)\times \vec{v}_{i}(t)}{|{\vec{r}}_{i,CM}(t)||\vec{v}_{i}(t)|} \right |,
\end{equation}
where $\vec{r}_{i,CM}(t) = \vec{r}_{i}(t) - \vec{R}_{CM}(t)$ and $\vec{R}_{CM}(t)$ is the center of mass (CM) position. 
 We use here a normalized sum of particles' angular momentum, a definition close to the one introduced by Erdmann et all\cite{erdmann2005noise}. In collective translation  $\varphi(t) \rightarrow 1$ and  $\Gamma(t) \rightarrow 0$. The opposite happens   in collective rotation, $\varphi(t) \rightarrow 0$ and  $\Gamma(t) \rightarrow 1$.

\section{Results}
\label{sec.results}

\subsection{Motion States}
\label{sec.motion}

We start varying the dimensionless parameters $Pe$, $\tau$ and $\beta$ at constant particle number, flexure number and null translational noise ($B = 0$). We measure  order parameters associated with the states of motion , Eqs.~\ref{9} and~\ref{10}. 

\begin{figure*}[!hbt]
\centering
\includegraphics[width=0.245\textwidth, height=4.27cm]{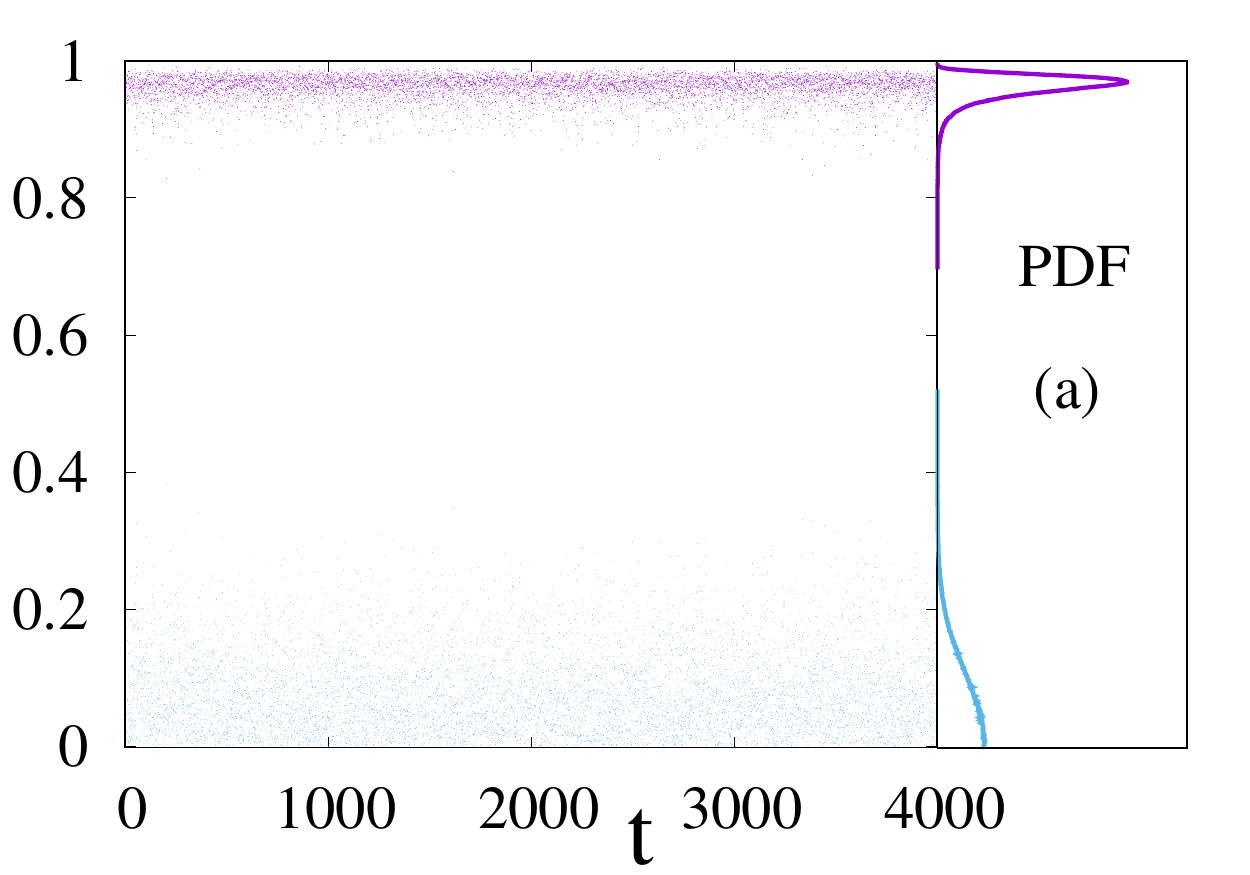}
\includegraphics[width=0.245\textwidth, height=4.27cm]{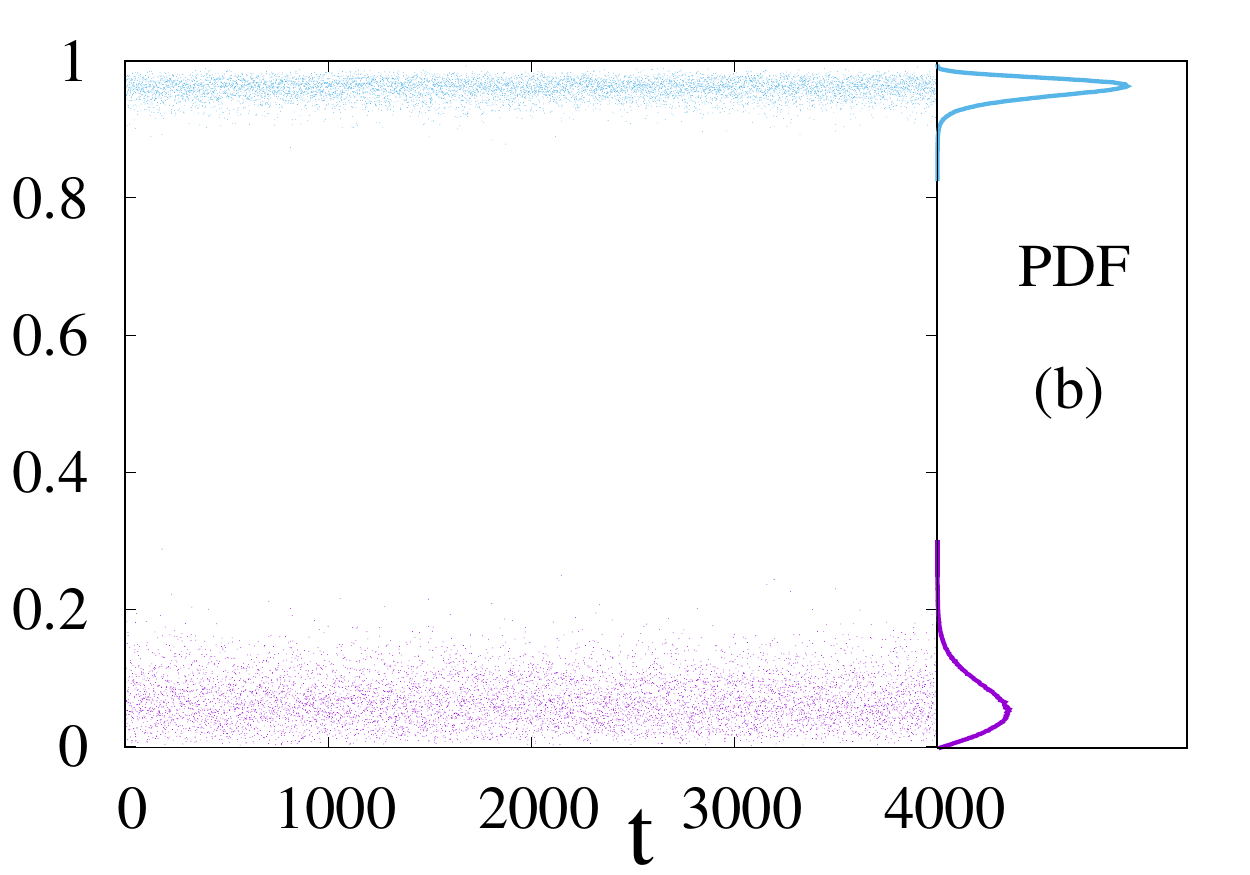}
\includegraphics[width=0.245\textwidth, height=4.27cm]{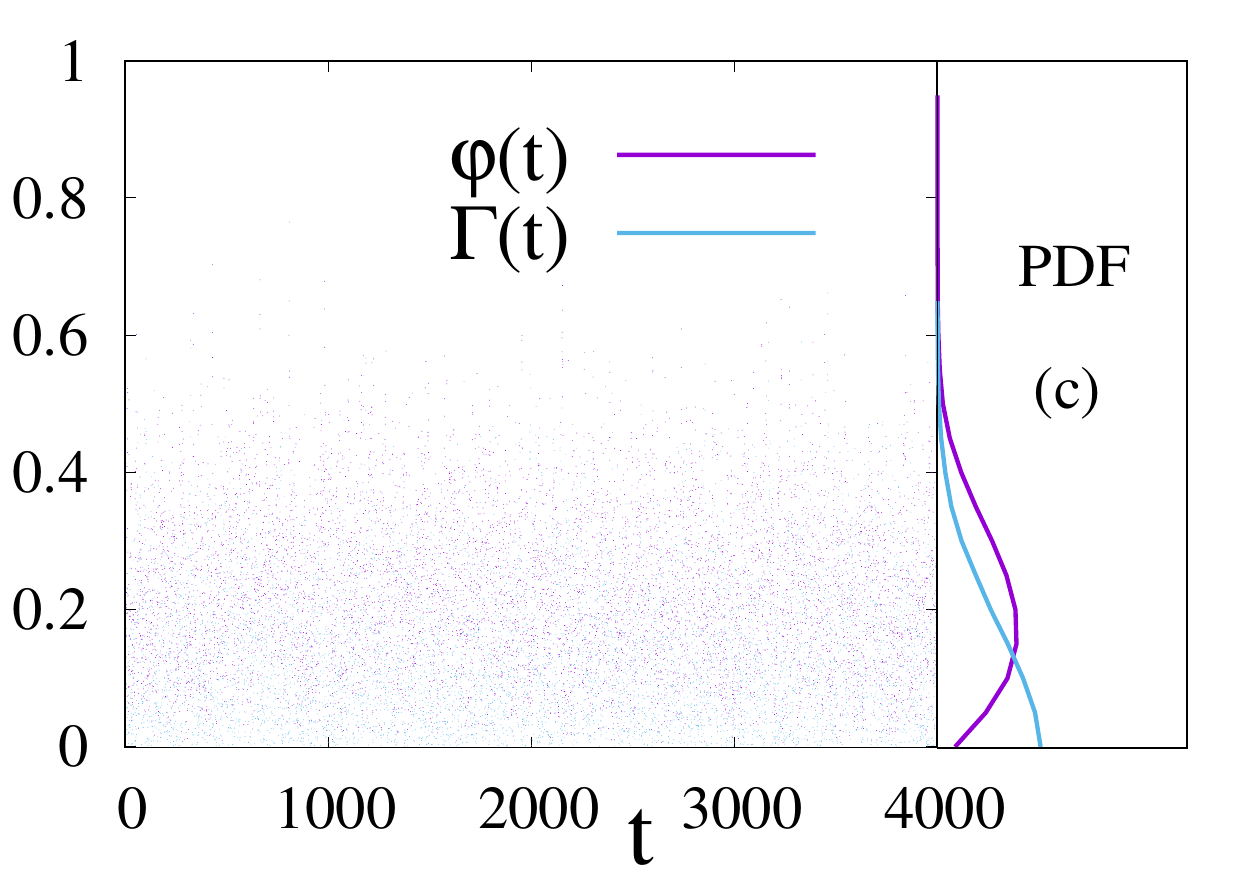}
\includegraphics[width=0.245\textwidth, height=4.27cm]{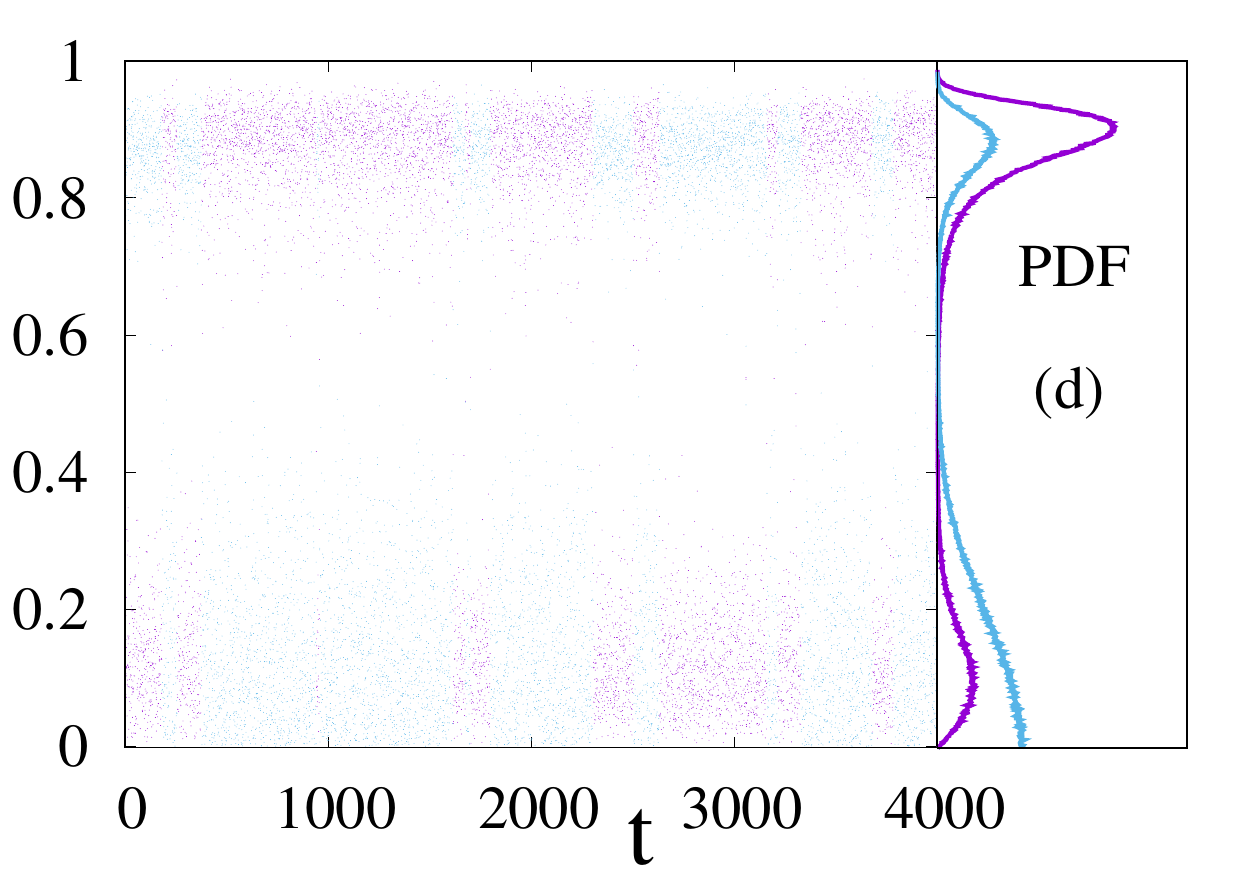}
\caption{
Steady state time series of the translational order parameter $\varphi(t)$ (purple) and  the rotational one $\Gamma(t)$ (blue) for a ring composed of $N = 20$ particles, no translational noise, $B = 0$, and flexure number $Fn = 1$. Corresponding probability distribution functions (PDF) for whole time series are also shown. Panels illustrate motions state observed in simulation for set of parameters $(Pe,\tau,\beta)$. (a) RUN state with (1,0.1,0), where fast relaxation of self-propelled velocity in the direction of velocity ensures a permanent translational collective motion with high persistence time. In this case, $\varphi(t)$ just fluctuates close to unity for all times. (b) ROT state with (1,0.1,1) in this case, $\Gamma(t)$ just fluctuate close to unity value for all times. (c) PRW state with (1,10,0), where rotational noise dominates angular dynamics causing all particles to behave as independent ones. (d) RRM state with (0.3,0.1,1); this case presents a dynamics where system alternates between RUN and ROT states; both order parameters alternate high and low values and a double PDF peak is observed. See Movies1-4 in Suplem. Material. See text for more details.
}
\label{serie-times}
\end{figure*}

For $N=20$, $\tau=0.1$ and $Pe= 1$, the ring reaches a stationary state induced by its initial configuration. That is, if $\beta = 0$, the system enters a translational collective (RUN) motion.  Figure~\ref{serie-times}a displays the evolution of the order parameters $\varphi$ and $\Gamma$. Figure~\ref{CM-XY}c illustrates the center of mass typical trajectory. When  $\beta = 1$, it enters a rotational (ROT) state. Figures~\ref{serie-times}b and~\ref{CM-XY}d show the order parameters and the center of mass trajectory, respectively.
For $Pe=1$ and  $\tau=10 $, the center of mass performs a persistent random walk (PRW). Fig.~\ref{serie-times}c illustrates this observation, and Fig.~\ref{CM-XY}a presents a center of mass trajectory for the same state, but for different parameters.
In an intermediary parameter region, $\tau=0.1$ and $Pe = 0.3$, the ring  switches between translation and rotation in a run and rotate mode (RRM).  This happens independently of the initial configurations. Figure~\ref{serie-times}d shows the time series and the probability distribution function correspondent to the state. In  Fig.~\ref{CM-XY}b,  we plot a typical center of mass trajectory. 

\begin{figure}[!h]
\centering
\includegraphics[width=1.0\columnwidth]{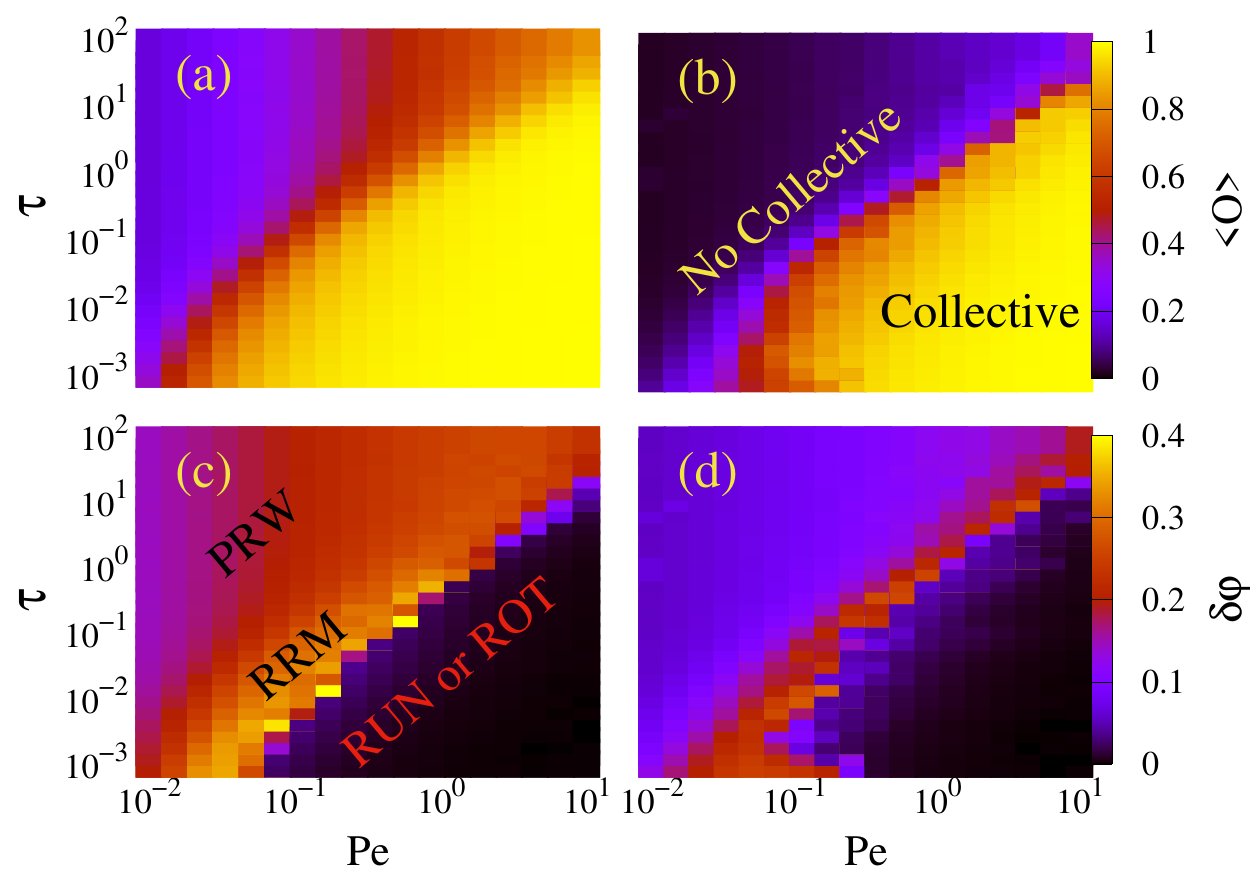}
\caption{
State diagram  ($Pe \times \tau$) for collective motion order parameter, $\left \langle O \right \rangle$, upper panels, and fluctuation of translational collective motion order parameter, $\delta \varphi$, bottom panels. Left panels show results for systems with $N  = 20$ and right ones for $N  = 200$, both use $Fn = 1$ and are initialize with $\beta = 0$. The PRW behavior of CM is found in the ABP limit, at high $\tau$ and no  collective motion. Single state collective motion settles for low values of $\tau$ and higher values of $Pe$, where both contribute to increase persistent motion of individual particles in such way that a mechanism similar to those of Ref.~\cite{szabo} is observed. ROT or RUN states are  obtained depending on initialization parameter $\beta$. At intermediary values of $\tau$ and $Pe$ parameters, we observe RRM motion: during a simulation system alternates between RUN and ROT states. This region, characterized by high fluctuation values, shrinks as ring's size increases and  disappears for large enough $N$.
Note that parameter axes are in log scale but the color bar is in linear scale.}
\label{phasediagram_collectivity_B0_beta0}
\end{figure}

 To characterize these different modes, we define an order parameter to identify collective motion regardless of its type:
\begin{equation}
\label{11}
    O(t) \equiv \varphi(t)^{2} + \Gamma(t)^{2} \,. 
\end{equation}
Collective motion is present when $O(t) \rightarrow 1$. Our systematic simulations resulted in the state diagram ($Pe \times \tau$) for two different initializations ($\beta = 0$ and $\beta = 1$). Fig. \ref{phasediagram_collectivity_B0_beta0} shows the case $\beta=0$, case $\beta=1$ is similar (not shown). 

So, independently of initialization, we observe in Fig. \ref{phasediagram_collectivity_B0_beta0}a-b a region where collective motion is settled ($\left \langle O \right \rangle \rightarrow 1$, yellow region) and another one with small values of  $\left \langle O \right \rangle$ (purple region) separated by an intermediary region (orange and red). Symbols $\left \langle . \right \rangle$ indicate time averages.

When $\tau/Pe \ll 1$,  we find $\left \langle O \right \rangle \rightarrow 1$. Meaning that, when the self-propelled velocity orientation  relaxation time, $\tau$, is much smaller than the rotational noise time scale,  $\tau_{R}$, particles quickly align their self-propelled velocities, $\hat{n}$, in the direction of the scattered velocity.
In this limit, and noting that each particle is always interacting with at least two neighboring particles, velocity alignment will occur according to the mechanism described in Ref.~\cite{szabo} and the initial condition determines whether the system will be in  rotational or translational collective motion, see Figs.~\ref{serie-times}a and~\ref{serie-times}b.

\begin{figure}[!h]
 \flushleft
    \includegraphics[width=1.0\columnwidth]{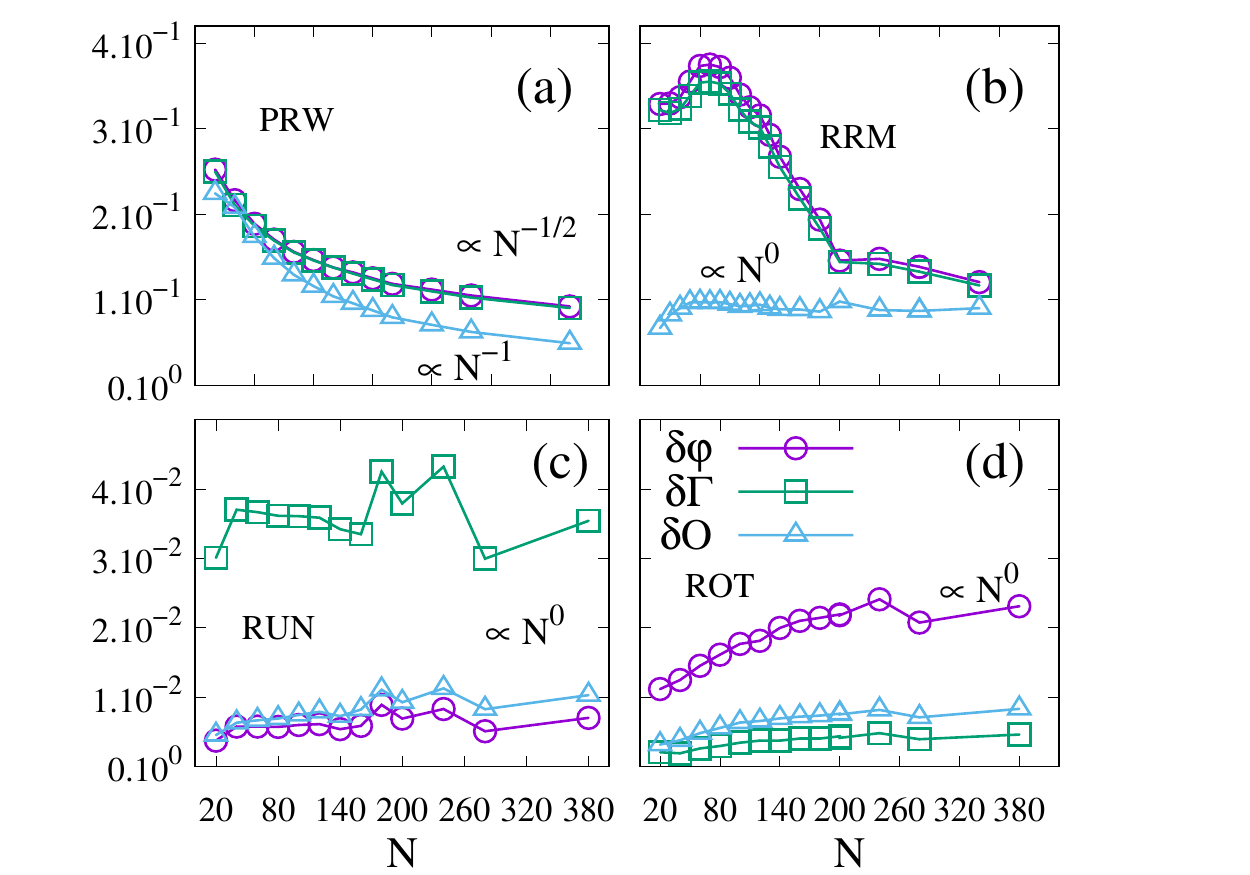}
\caption{
Fluctuation of different collective motion order parameters as functions of $N$ for a set of parameters $(Pe,\tau,\beta)$ with fixed $B = 0$ and $Fn = 1$: (a) Region of absence of collective motion in the ABP limit, $\tau/Pe \gg 1$, for parameters (1,100,0). In this region, the system presents a persistent random walk for the CM movement, resulting from the summation of random variables, 
where fluctuations are expect to decrease with $N^{-1/2}$. (b) Region RRM, (0.25,0.1,0), where after an initial increase with $N$ it gets more difficult to system to change between RUN and ROT states. 
(c) and (d) regions of translational, (4,0.1,0), and rotational collective motion, (4,0.1,1), respectively. In both regions, fluctuations are almost independent of $N$ since all particles spend most time completely aligned in a translational motion or rotating around CM.
}
\label{fluctuations_B0_beta0}
\end{figure}

On the other side, when $\tau/Pe\gg 1$, angular noise destroys collective motion implying that $\left \langle O \right \rangle \rightarrow 0$.
Self-propelled velocities of different particles become uncorrelated, the center of mass motion resulting from a sum of random displacements.
 This limit  corresponds to the case of active Brownian particles (ABP), a prototypical model to study competition between noise and self-propulsion effects~\cite{schweitzer2003brownian,romanczuk2012active,digregorio2018}.
In a ring with a small number of particles (Fig.~\ref{phasediagram_collectivity_B0_beta0}a and Fig.~\ref{phasediagram_collectivity_B0_beta0}c), we observe a mean value above zero  (purple color) for $O(t)$ in the disordered state, this happens because fluctuations $\delta\phi$ and $\delta \Gamma $ of both order parameters scale with $1/\sqrt{N}$, so fluctuations in $O(t)$ scale with $1/{N}$ (Fig.~\ref{fluctuations_B0_beta0}a).
In the Supp. Material C, we analyze the dependence of $\delta \varphi$ with $\sqrt{N}$  in the ABP limit ($ \tau/ Pe \gg  1$). For small number of particles, fluctuations are high in the RW region. For large particle numbers, the system reaches a disordered state with small fluctuations, as shown in  Fig. \ref{phasediagram_collectivity_B0_beta0}b, Fig. \ref{phasediagram_collectivity_B0_beta0}d and Fig. \ref{fluctuations_B0_beta0}a.

 In the intermediary region which separates  disordered and ordered  states,  $\tau/Pe \sim 1$,
 fluctuations of both order parameter, $\varphi(t)$ and $\Gamma(t)$, increase indicating the emergence of a distinct motion state where the ring switches between rotation and translation. We call this run and rotate motion (RRM) and show a typical time series in Fig.~\ref{serie-times}d. 
   Since parameters $\varphi$ and $\Gamma$ are complementary in the RRM state, we use their  fluctuations $\delta \varphi$ and $\delta \Gamma$, to study it. When both fluctuations are close to zero, the system is out of the RRM state, being either in  collective or in persistent random walk states. 
 Comparing Fig.~\ref{phasediagram_collectivity_B0_beta0}c and Fig.~\ref{phasediagram_collectivity_B0_beta0}d, we observe the shrinking of the RRM region as the ring particles' number increases from $N=20$ to $N=200$. The decay in the fluctuations $\delta\phi$ and $\delta\Gamma$ (
 Fig.~\ref{fluctuations_B0_beta0}b) with $N$  confirms this tendency: they decrease up to $N=200$, remaining nearly constant for larger $N$ values. Finally, it is interesting to note in
Figures~\ref{fluctuations_B0_beta0}c and~\ref{fluctuations_B0_beta0}d 
that the fluctuations scale as $N^0$. In both cases, the system presents collective motion. For comparison, we use the same scales of Figures~\ref{fluctuations_B0_beta0}a and~\ref{fluctuations_B0_beta0}b. 

\subsection{Diffusion: Mean Square Displacement (MSD)} 
\label{sec.msd}

\begin{figure}[h]
\centering
\includegraphics[width=1.0\columnwidth]{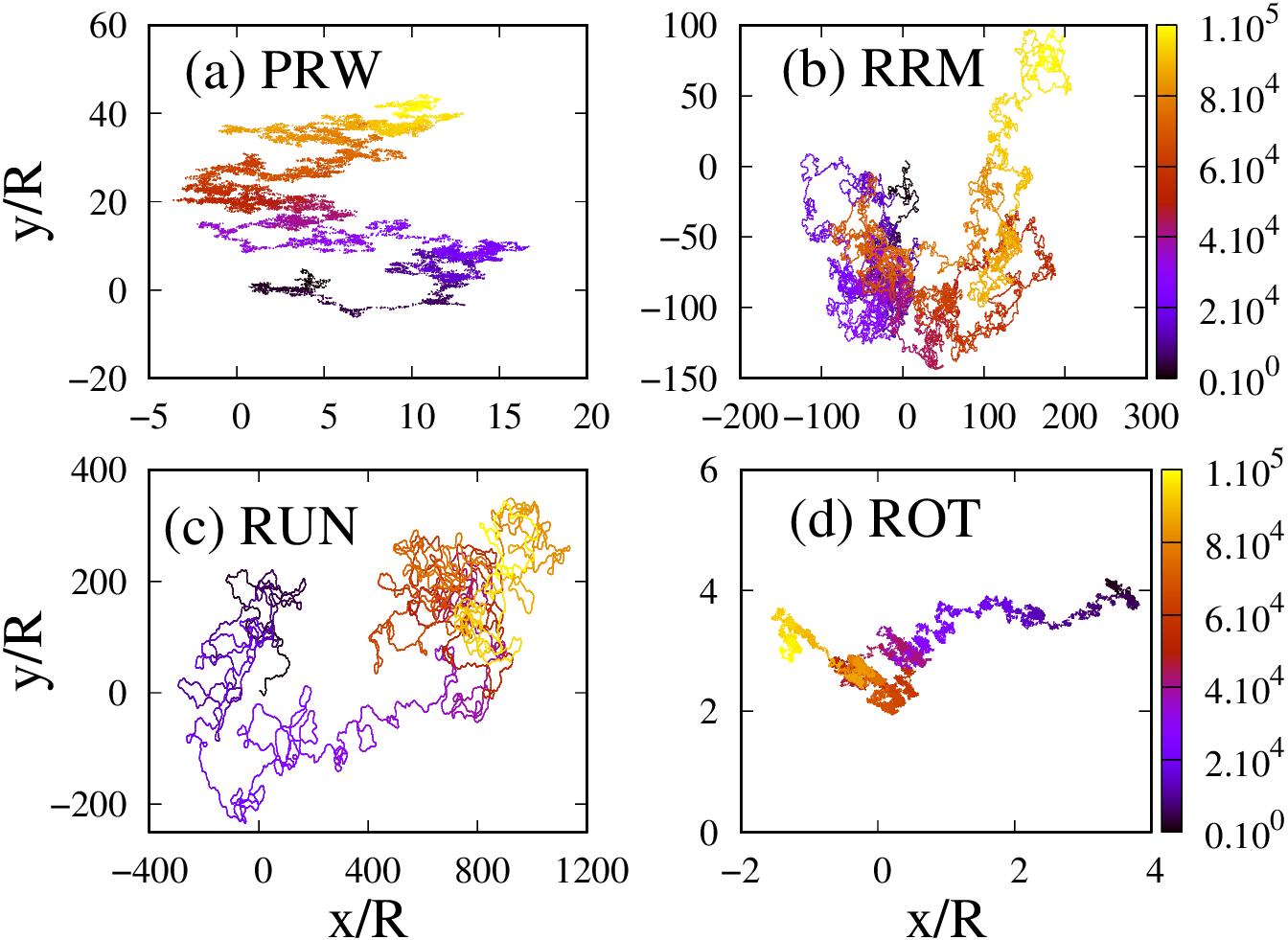}
\caption{
Typical CM evolution in a xy plot for a set of parameters $(Pe,\tau,\beta)$, with $N = 20$, $B = 0$ and $Fn = 1$  kept fixed: (a) PRW motion, (0.1,0.3,0); (b) RRM motion, (0.3,0.1,0); (c) RUN motion, (2,0.1,0); and (d) ROT motion, (2,0.1,1). Note that all panels have different xy ranges and that RUN motion presents the wider displacement region due to persistent collective motion of the CM. ROT state show very small displacement since particles rotate around CM. RRM presents intermediate values of displacements since it depends, essentially, in the fraction of time spend in RUN state. RW motion presents the expected behaviour of persistent motion of ABP particles where there is no collective motion.
Colors indicate time (see color bar).}
\label{CM-XY}
\end{figure}

\begin{figure}[!h]
\centering
\includegraphics[width=4.35cm, height=4.27cm]{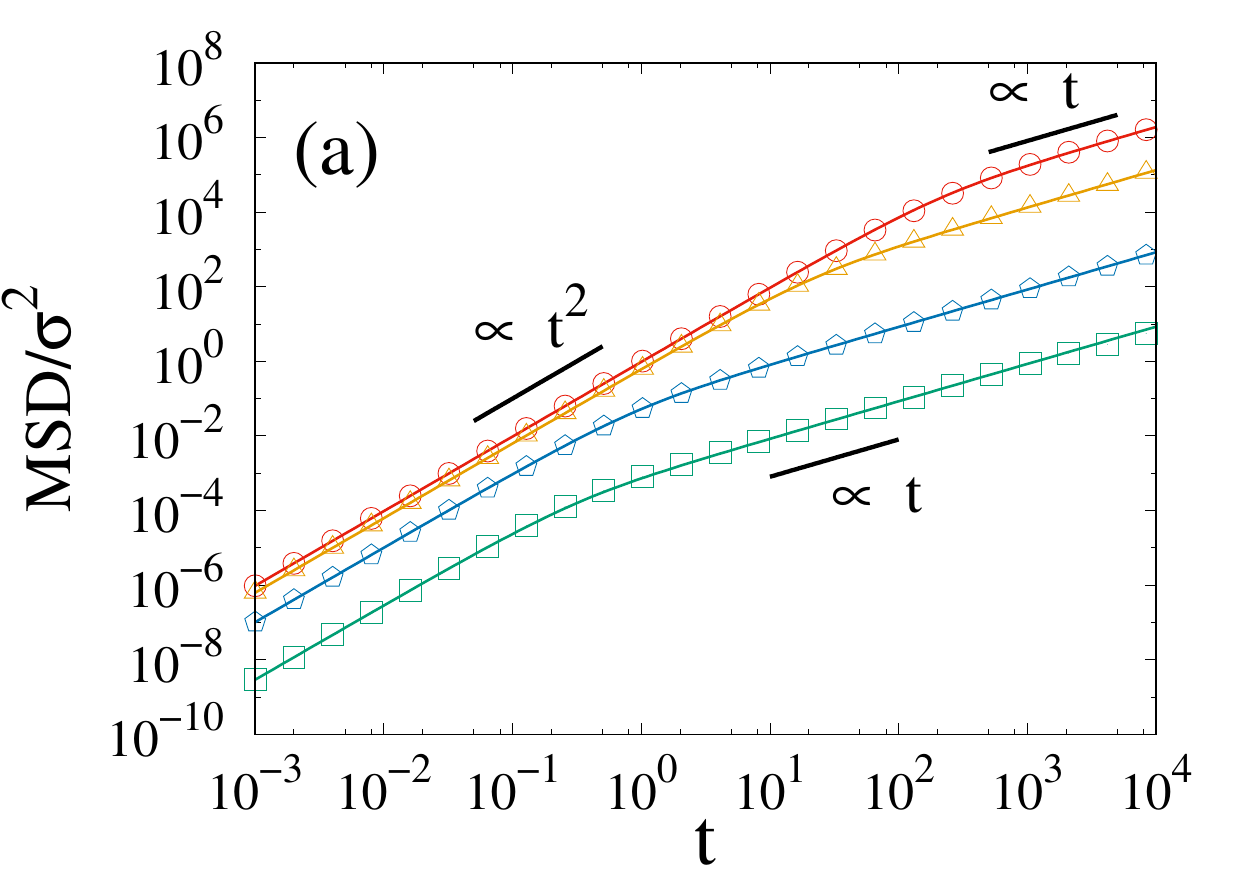}
\includegraphics[width=4.0cm, height=4.27cm]{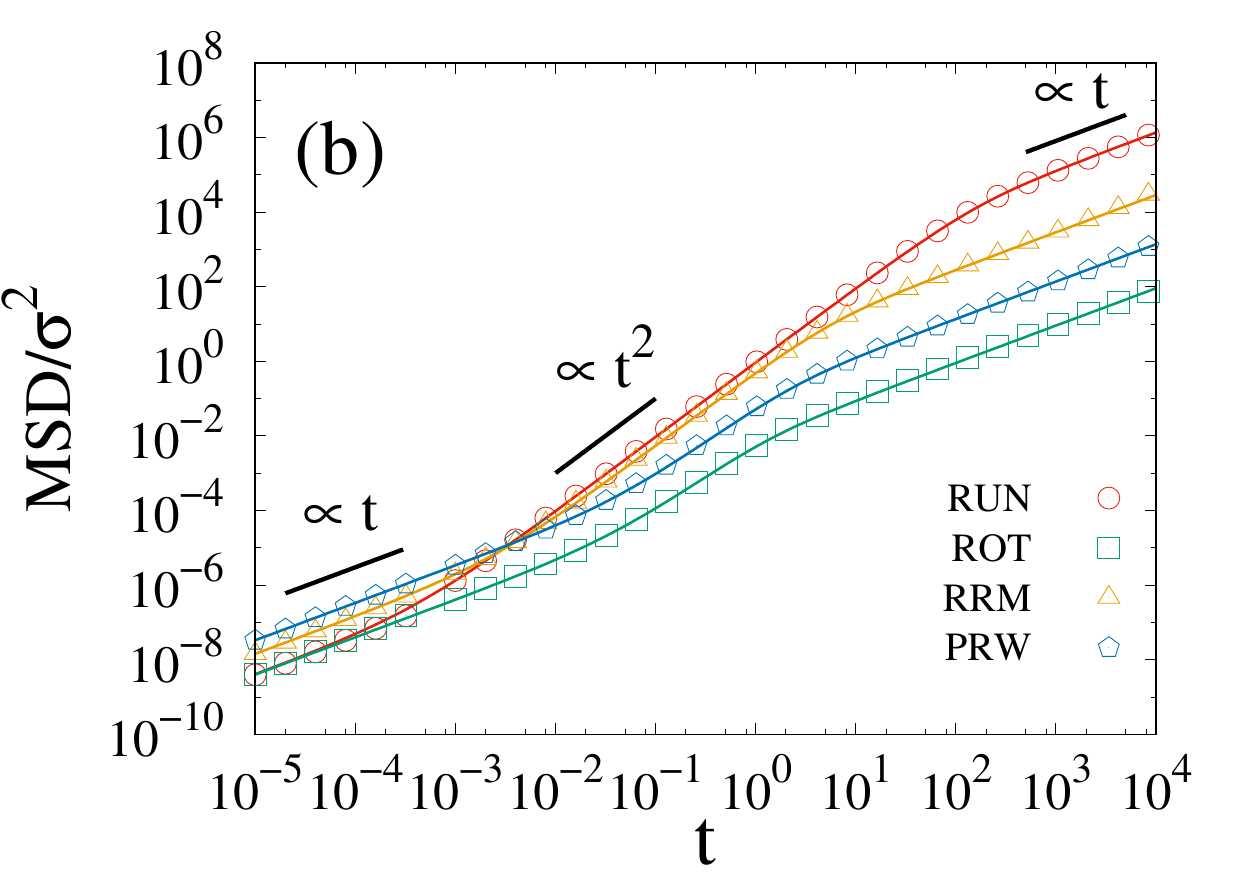}
\caption{
Center of Mass Mean Square Displacement ($MSD$) for the same parameters of Fig.~\ref{CM-XY} without (panel (a)), and with (panel (b)) translational noise. Note, however, that for $MSD$ estimations, we also perform averages over repeated simulations with different noise. In both panels, solid lines correspond to fit of Eq.~\ref{msd}, while the dots indicate numerical results. Fitted parameters are shown in Table~\ref{tab:table1}. RUN state exhibits highest values of persistent time and diffusion coefficient, followed by RRM state and RW. In ROT state, particles spend most of time rotating around the CM implying no significant displacement of it. When translational noise is present, $B = 0.001$, ballistic regime is preceded by a diffusive one.
}
\label{MSD-states}
\end{figure}

In this section, we characterize how the ring's center of mass behaves in the different motion states. To illustrate, in
 Fig.~\ref{CM-XY} we show typical center of mass trajectories for each motion state. Note the difference in scales in each case. The active ring has a longer reach when in RUN state (Fig.~\ref{CM-XY}c). The ring diffusion is characterized by the center of mass mean-square displacement, $MSD$, obtained by the sliding windowing method. In addition, we take averages over 40 trajectories\cite{Gal2013}. The correspondent center of mass $MSD$ (Figure~\ref{MSD-states}a, red curve) shows a long time interval in the ballistic ($\propto t^2$) regime. In Fig.~\ref{MSD-states}b we find a similar behavior, but a short time diffusive regime appears due to translational noise. 
 In both cases, the behavior is diffusive for asymptotically large times, as expected. RRM state (Figure~\ref{CM-XY}b and yellow curves in Figures~\ref{MSD-states}a,b), PRW state (Figure~\ref{CM-XY}a and blue curves in Figures~\ref{MSD-states}a,b) and ROT state (Figure~\ref{CM-XY}d and green curves in Figures~\ref{MSD-states}a,b) present similar trends, but successively smaller ballistic regimes, implying smaller asymptotic diffusion constants.
 
\begin{figure}[!h]
\centering
\includegraphics[width=4.35cm, height=4.27cm]{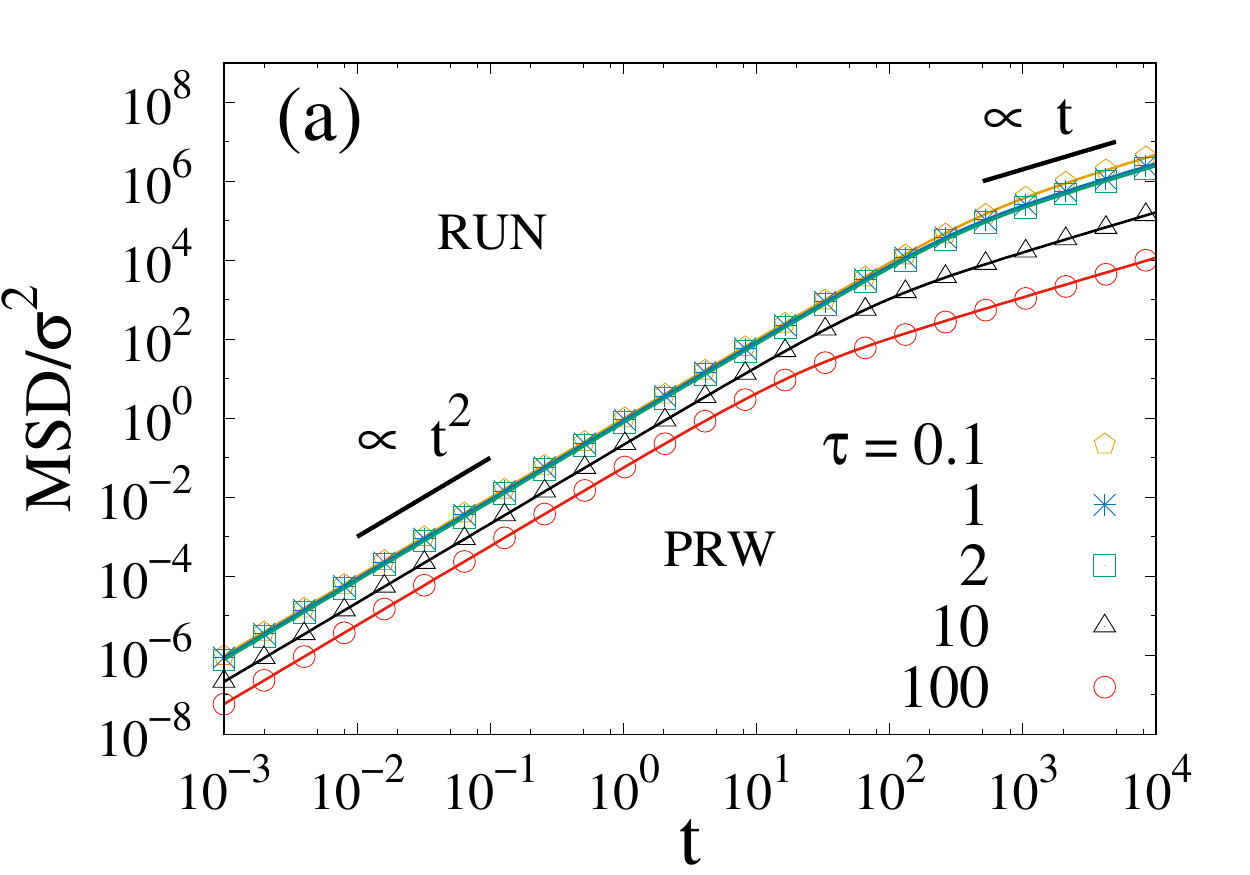}
\includegraphics[width=4.2cm, height=4.27cm]{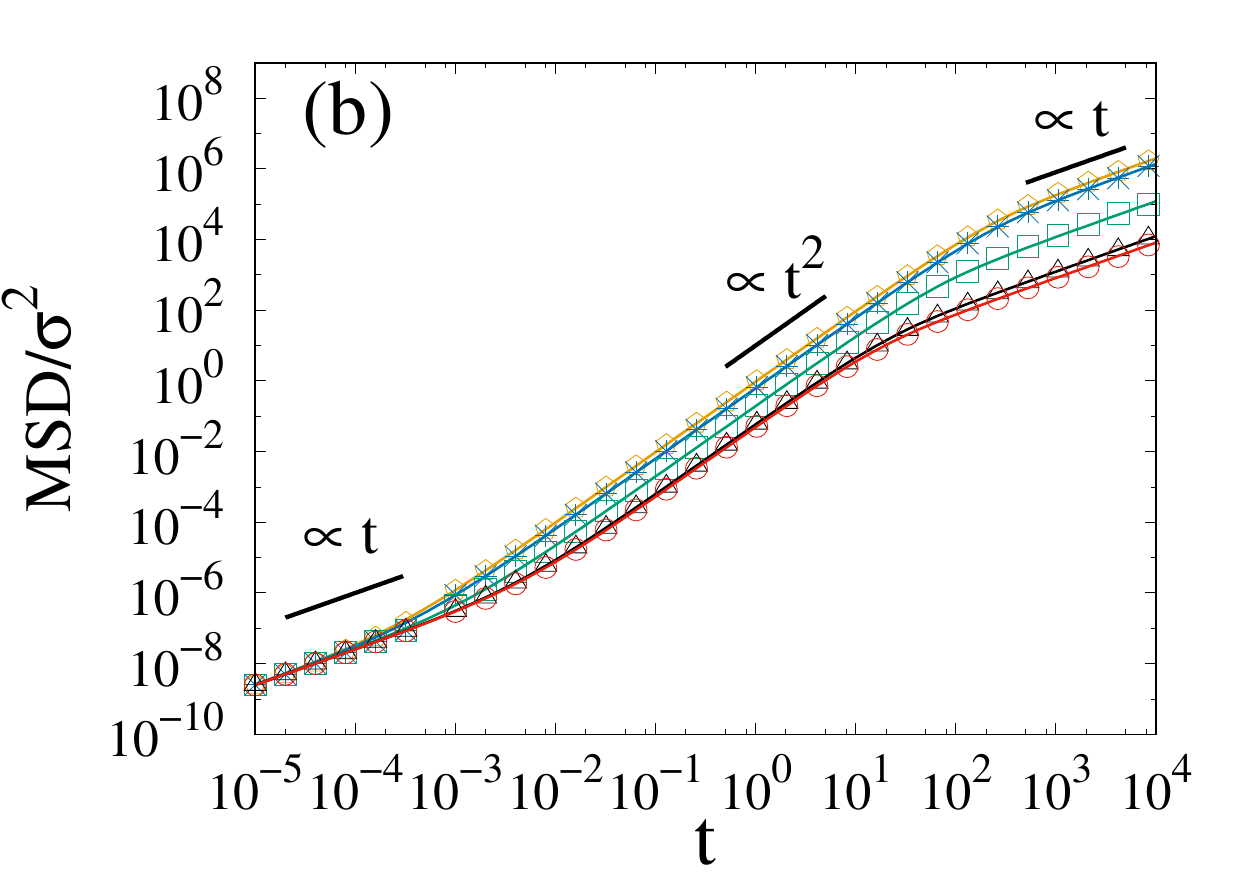}
\includegraphics[width=4.27cm, height=4.27cm]{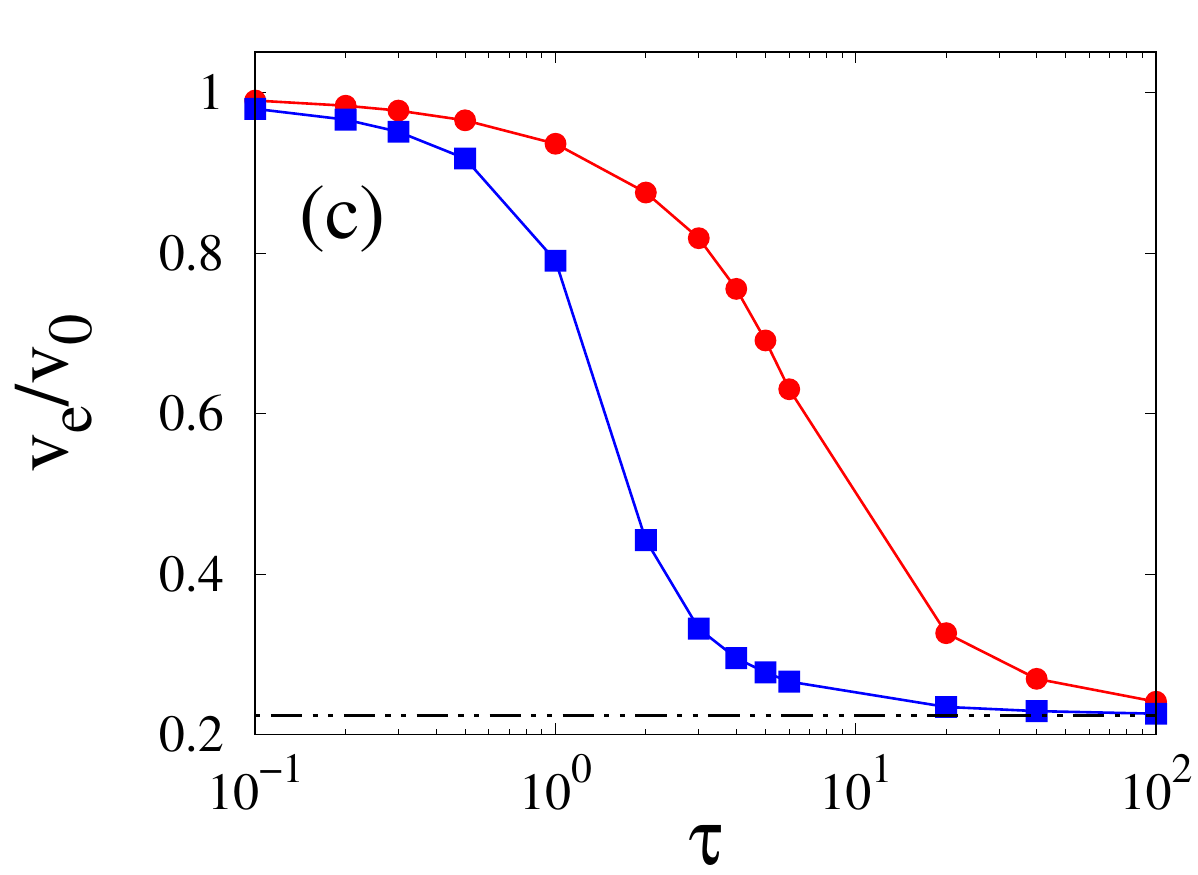}
\includegraphics[width=4.27cm, height=4.27cm]{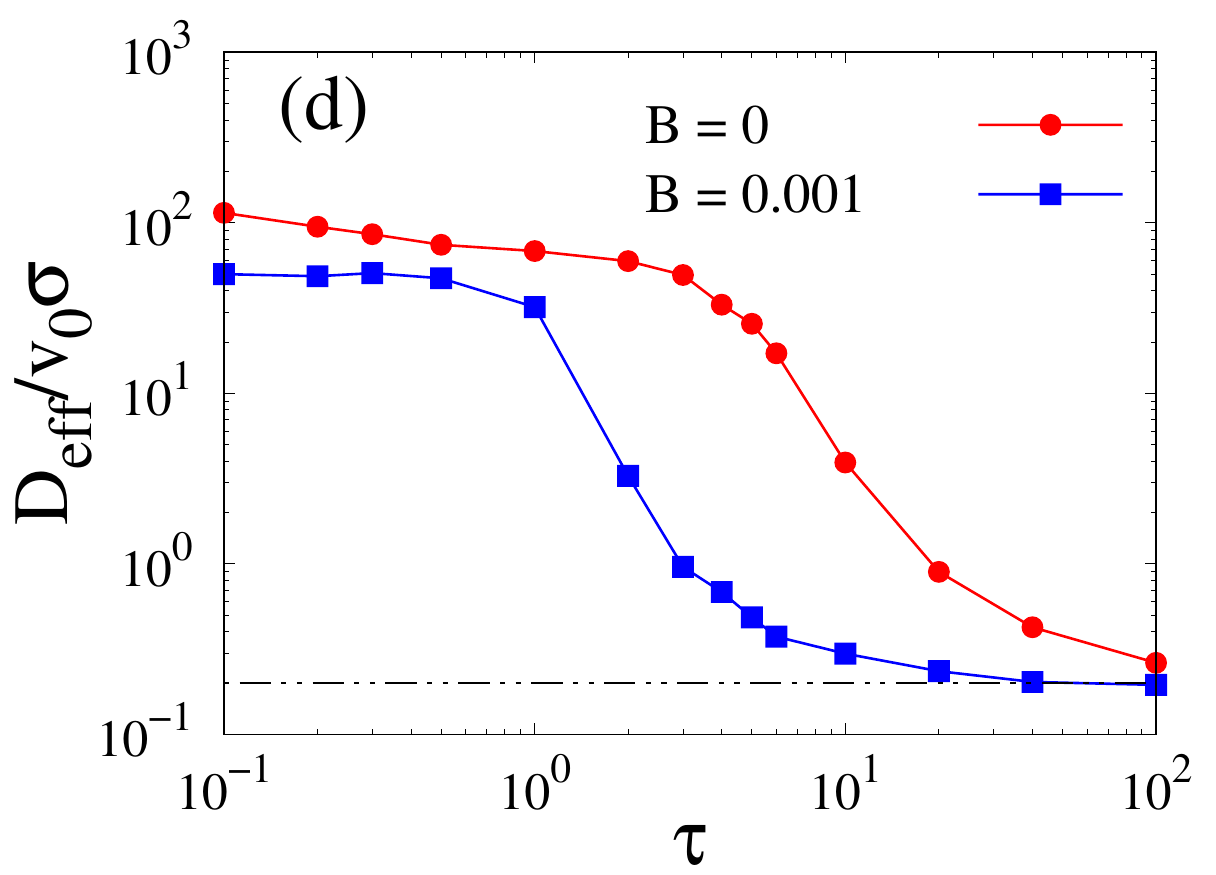}
\caption{
Center of Mass Mean Square Displacement ($MSD$) for systems without (panel (a)), and with translational noise ($B = 0.001$ in Panel (b)). Parameters $N = 20$, $Pe = 4$, $\beta = 0$ and $Fn = 1$ are kept fixed. Both figures show different values of  parameter $\tau$ from high values of it, corresponding to RW state (ABP limit and absence of collective motion), to low values, corresponding to RUN state, where all particles move aligned to the CM velocity. Solid lines correspond to fit of Eq.~\ref{msd}, while dots indicate numerical results. (c) Dependence of fitted effective self-propelled speed $v_{e}$ with $\tau$: in RUN state, all particles are aligned and move with velocities close to $v_0$; in PRW state, it approaches theoretical predicted value (dashed line). (d) Fitted effective constant diffusion as a function of $\tau$: note that the transition at low $\tau$ values. Note, also, that even a small translational noise (panels (c) and (d)) is sufficient to displace the position of transition to collective motion by an order of magnitude in $\tau$. 
}
\label{MSD_Pe4_N20}
\end{figure}

In the limit $\tau/Pe\gg 1$, each particle behaves as an ABP~\cite{romanczuk2012active,basu2018active,bechinger2016active,fodor2018statistical,schweitzer2003brownian}, and the ring as a whole executes a PRW.
 In this limit, there is an analytical solution for the $MSD$~\cite{schweitzer2003brownian,bechinger2016active}, which we detail in Supp. Material A and B. 
 In Fig.~\ref{MSD_Pe4_N20}a we show the $MSD$ obtained for the active ring for different relaxation time values, $\tau$, fixed $Pe = 4$ and $\beta=0$. In Fig.~\ref{MSD_Pe4_N20}b we use the same parameters and include a small translational noise ($B=0.001$).  
In both cases, increase in $\tau$ reduces the ballistic region extension resulting in a smaller long time diffusion.
We note that behavior shown in Figs.~\ref{MSD-states} and~\ref{MSD_Pe4_N20} qualitatively resembles the exact solution for ABPs. This observation suggests the possibility of  fitting effective parameters combining translational and rotational noise time scales,
\begin{equation}
\label{msd}
MSD(t) = \frac{4\sigma^2}{N\tau_T} \, t\, 
+ 2(v_e\tau_e)^{2} \, \left [\frac{t}{\tau_{e}} + (e^{-t/\tau_{e}}-1) \right ],
\end{equation}
where $\tau_{e}$ and $v_{e}$  are the effective persistence time and the effective self-propelled speed, respectively.  
 In Supp. Mat. B, we detail the relation among  $v_{e}$ and the center of mass mean square velocity. 
At long times, $t\gg \tau_e$, we find 
\begin{equation}
\label{diff_constant_effect}
     D_{eff} = \lim_{t \gg \tau_{e}}  \frac{MSD(t)}{4t} = \frac{\sigma^2}{N\tau_T} +\frac{v_{e}^{2}\tau_{e}}{2}.
\end{equation}
\begin{table*}[!th]
 \caption{
 Parameters describing curves of Fig.~\ref{MSD-states} for fixed parameters $N = 20$ and  $Fn = 1$ are shown. Effective persistence times and self-propelled velocities are obtained by fitting Eq.~[\ref{msd}] to the MSD simulation's data. For comparison we show several ratios of these fitted quantities with other system parameters. Fig.~\ref{CM-XY}  shows the trajectories of systems in the absence of translational noise.  
 }
 \label{tab:table1}
 \begin{tabular}{p{0.25\linewidth}p{0.04\linewidth}p{0.04\linewidth}p{0.04\linewidth}p{0.08\linewidth}p{0.08\linewidth}p{0.08\linewidth}p{0.08\linewidth}p{0.08\linewidth}}
    \hline
   State of Motion & $Pe$ & $\tau$ & $\beta$& $B$ &$\tau_{e}/\tau_{0}$  & $\tau_{e}/\tau_{R}$ & $v_{e}/v_{0}$ & $D_{eff}/v_{0}\sigma$  \\
    \hline
      Pers. Random Walk     & 0.1 & 0.3 & 0  & 0      & 0.41   & 2.05  & 0.32   & 0.02 \\ 
      Pers. Random Walk     & 0.3 & 0.1 & 0  & 0.001  & 1.04   & 1.73  & 0.275  & 0.034\\
      Run and Rotate  & 0.3 & 0.1 & 0  & 0      & 10.6   & 17.68 & 0.786  & 3.15 \\ 
      Run and Rotate  & 0.7 & 0.1 & 0  & 0.001  & 2.6    & 1.86  & 0.74   & 0.76 \\ 
      Run             & 2   & 0.1 & 0  & 0      & 99.32  & 24.83 & 0.98   & 47.6  \\ 
      Run             & 2.5 & 0.1 & 0  & 0.001  & 73.05  & 14.61 & 0.96   & 33.02 \\ 
      Rotate          & 2   & 0.1 & 1  & 0      & 0.144  & 0.036 & 0.054  & 0.000169\\
      Rotate          & 2.5 & 0.1 & 1  & 0.001  & 0.55       & 0.11  & 0.089  & 0.00228\\ \hline
  \end{tabular}
\end{table*}
Table~\ref{tab:table1} displays fitted parameters for the center of mass $MSD$, Eq.~\ref{msd}, of displacements of Figs.~\ref{CM-XY}. Note the large increase in the effective persistence time in the RUN state.
In Fig.~\ref{MSD_Pe4_N20}c and Fig.~\ref{MSD_Pe4_N20}d, we also show the effective values for velocities and diffusion coefficients obtained fitting simulation data with that equation. Note the transition  between the low-$\tau/Pe$, collective motion state,  and high-$\tau/Pe$, ABP state.

\subsubsection{High-$\tau/Pe$ limit}
In the ABP limit, or high-$\tau/Pe$ limit, the first term in Eq.~\ref{2} is negligible and the rotational dynamics is dominated by rotational noise, that is, for $N=20$,
$$\tau_{e} \rightarrow  \tau_R\;, \frac{v_{e}}{v_0} = \frac{1}{\sqrt{N}}\sim0.22\;,$$
as seen in Fig~\ref{MSD_Pe4_N20}c.
In this limit, Eq.~\ref{diff_constant_effect}  results in
\begin{equation}
    \frac{D_{eff}}{v_0\sigma}=\frac{B}{2{N}Pe}+\frac{Pe}{N}\sim0.2\;,
    \label{diff-high-tau}
\end{equation}
 as can be checked in Fig.~\ref{MSD_Pe4_N20}d for $N=20$. The first term in Eq.~\ref{diff-high-tau} is negligible, since $B=0.001$. As $\tau$ decreases, both $v_e$ and $D_{eff}$ depart from the ABP  behavior (Figure~\ref{MSD_Pe4_N20}c). Nevertheless, note that even a small translational noise shifts this departure by one order of magnitude (Figure~\ref{MSD_Pe4_N20}d). 

\subsubsection{Low-$\tau/Pe$ limit} 

The ballistic regime is extended in this limit,
 resulting in large effective persistence times (Figures~\ref{MSD_Pe4_N20}a and~\ref{MSD_Pe4_N20}b). Also, the effective velocity reaches $v_0$ (Figure~\ref{MSD_Pe4_N20}c), so the whole ring achieves the free single-particle speed. The consequence is a high diffusion constant, as can be checked in Fig.~\ref{MSD_Pe4_N20}d.


\subsubsection{Diffusion and Ring Size}

\begin{figure}[!htb]
\centering
\includegraphics[width=4.27cm, height=4.27cm]{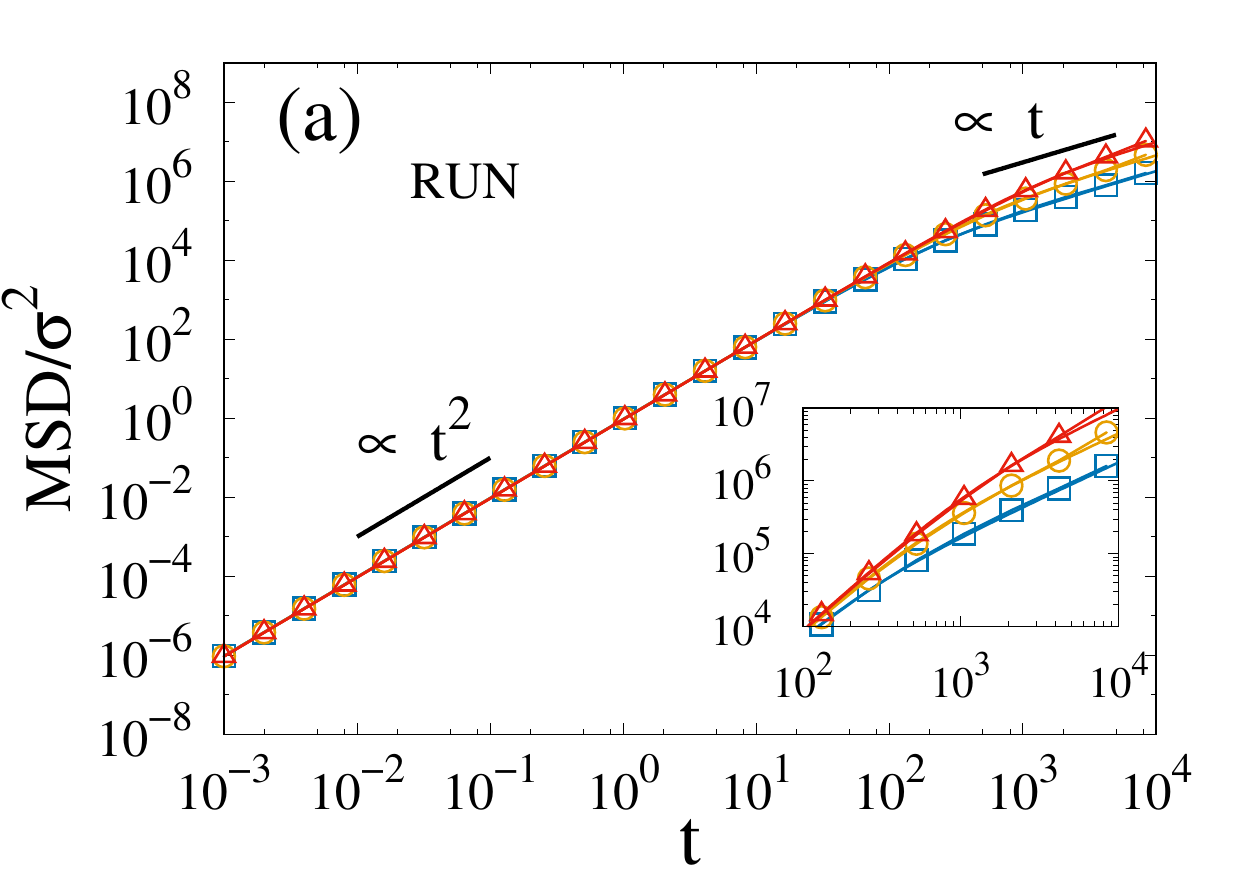} 
\includegraphics[width=4.27cm, height=4.27cm]{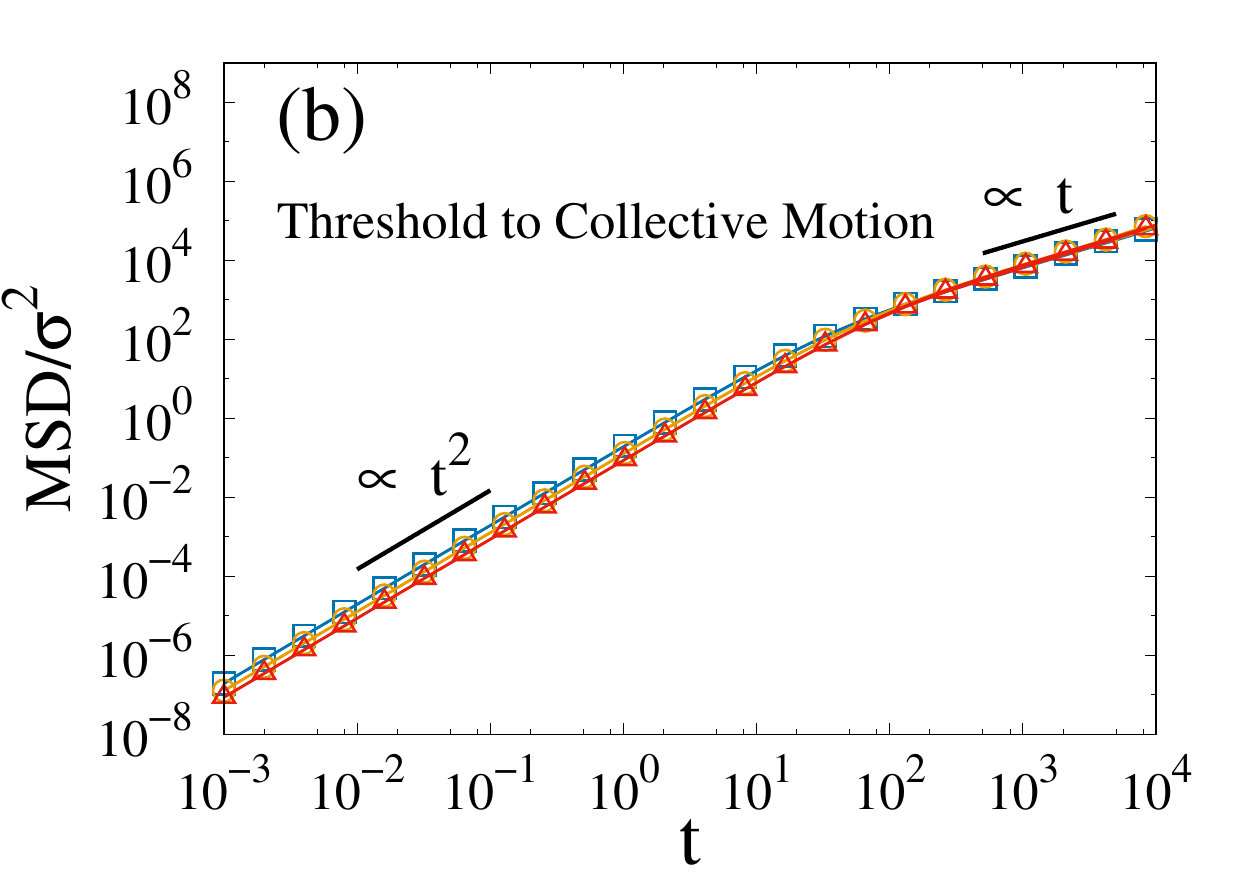} 
\includegraphics[width=4.27cm, height=4.27cm]{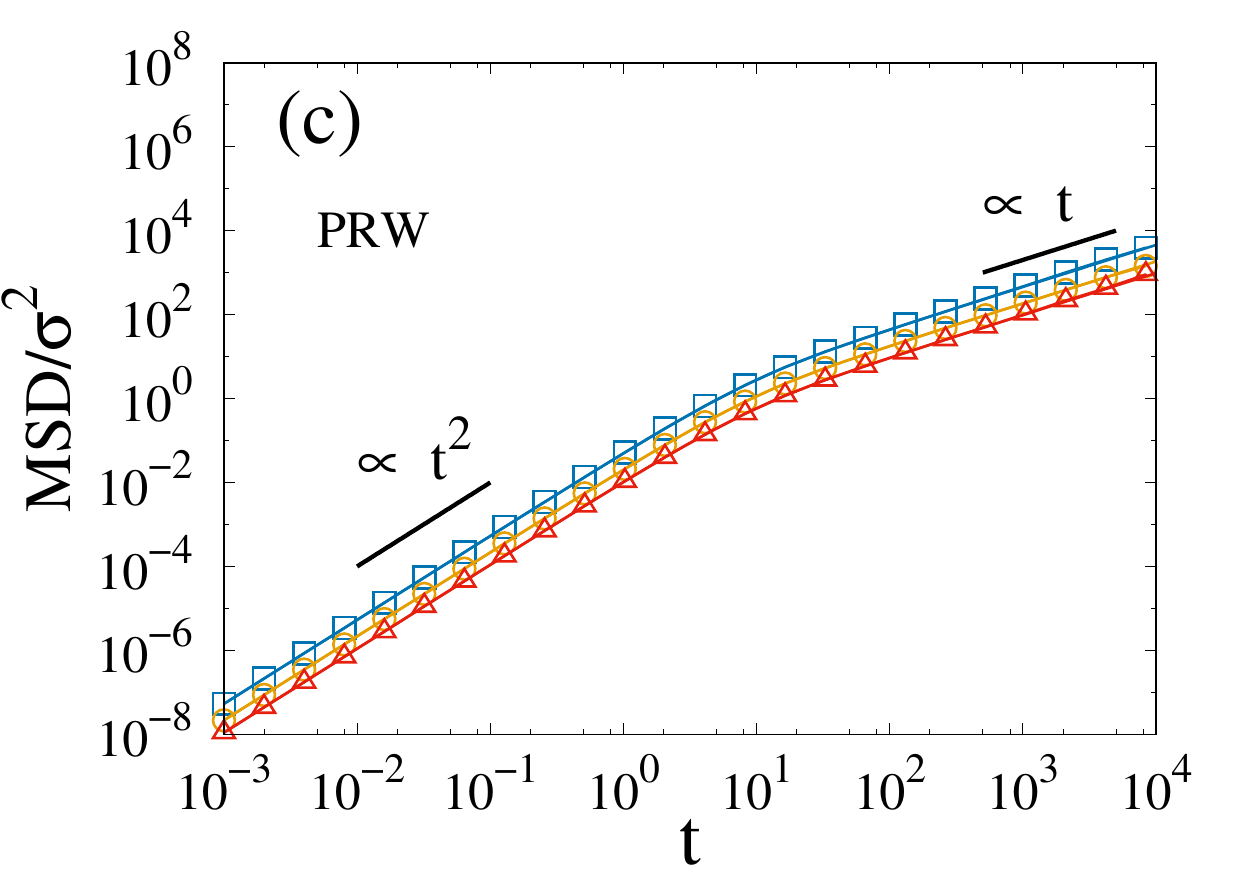} 
\includegraphics[width=4.27cm, height=4.27cm]{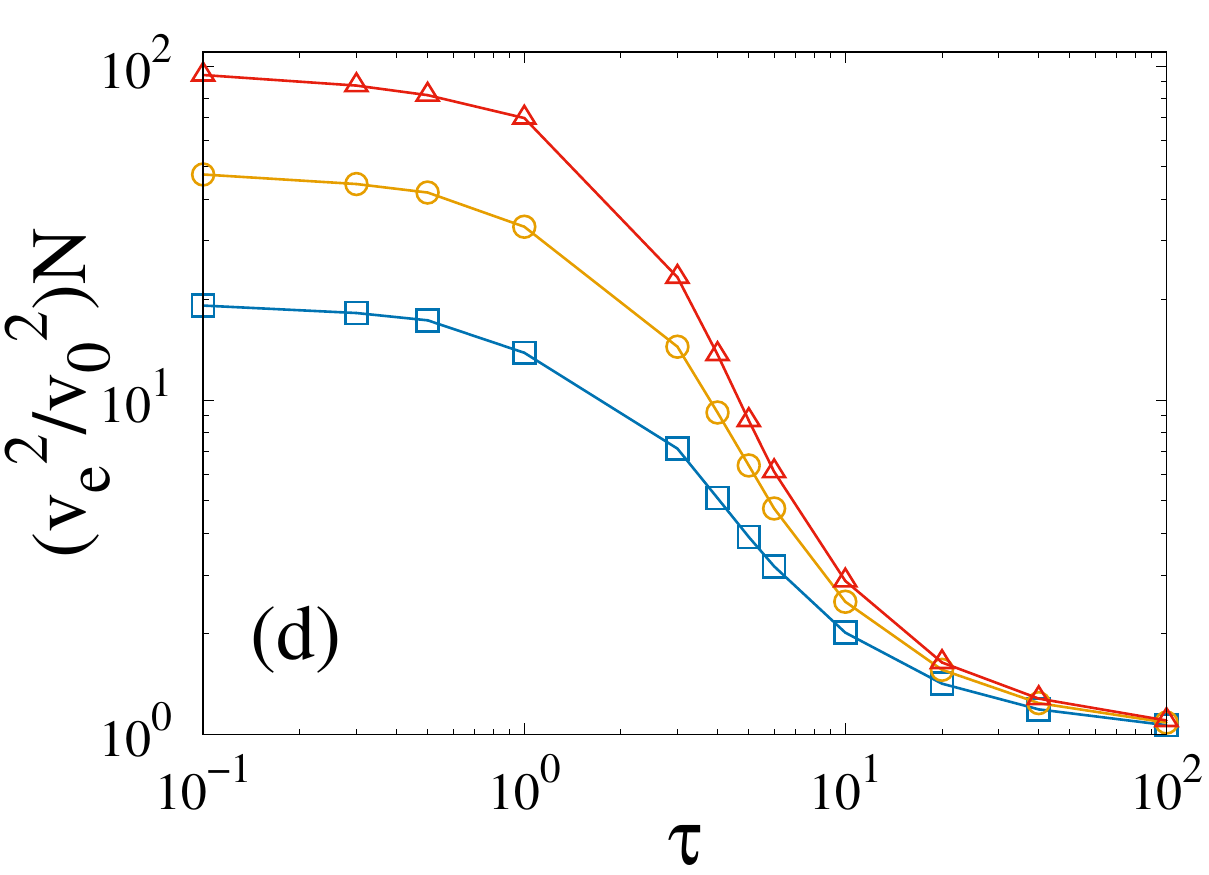}
\includegraphics[width=4.27cm, height=4.27cm]{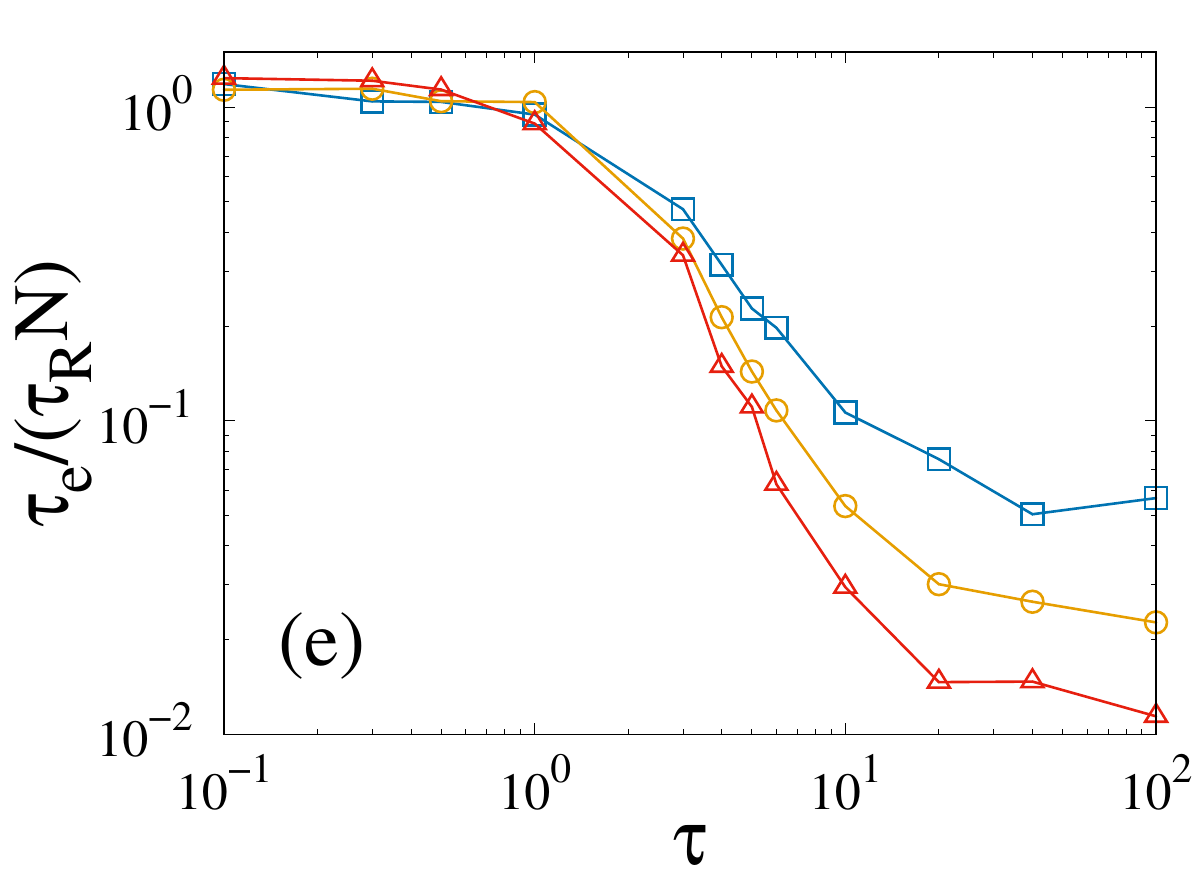}
\includegraphics[width=4.27cm, height=4.27cm]{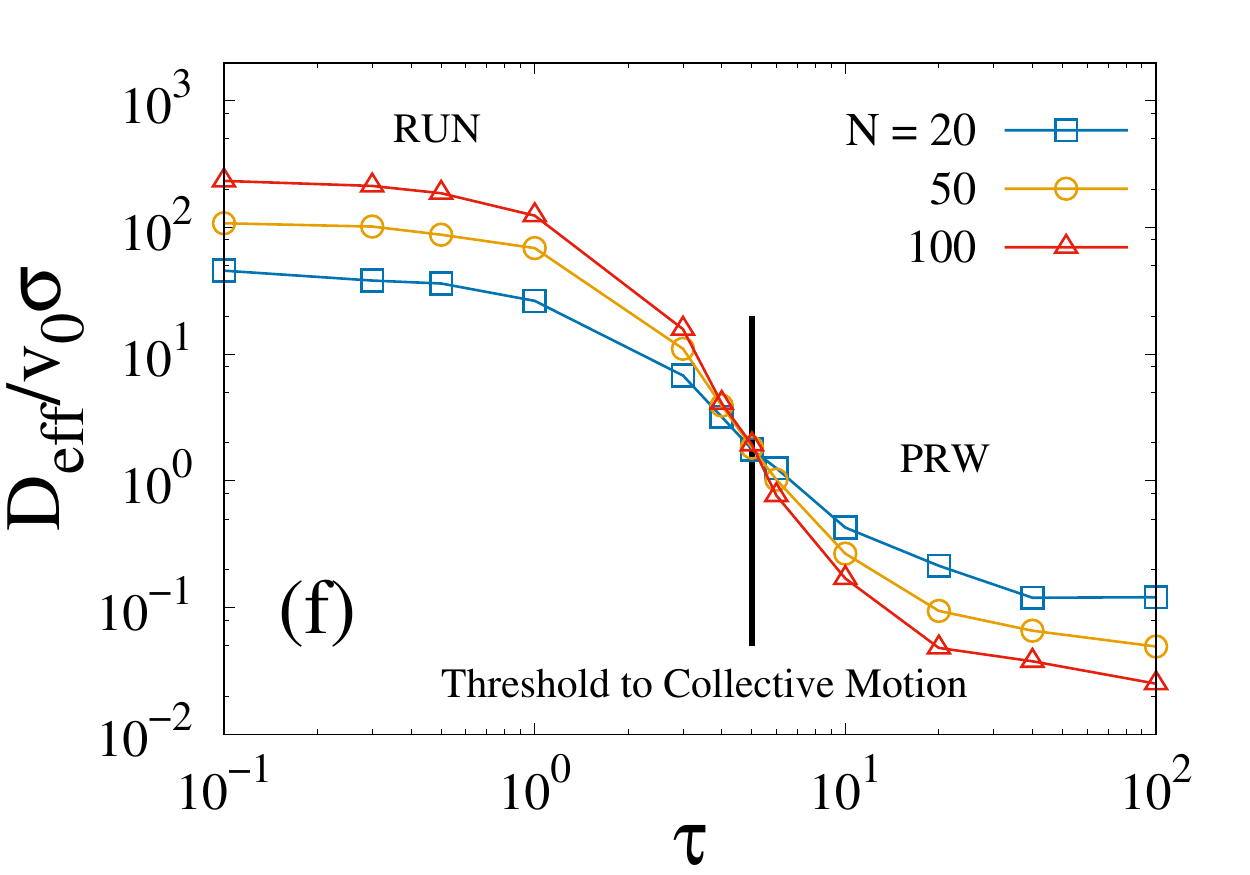} 
\caption{
First panels show the change in behavior of $MSD$ of Center of Mass as system goes from (a) RUN state, $\tau = 0.1$, to (c) PRW state, $\tau = 100$, crossing the (b) threshold region, $\tau = 5$, for fixed set of parameters: $Pe = 2$, $\beta = 0$, $B = 0$ and $Fn = 1$. We observe that in the RUN state, larger systems present higher values of effective diffusion constant and persist longer in ballistic regime (inset). In the threshold region, diffusive behavior is independent of system's size. PRW region shows the expected decrease of diffusion constant with $N$. Panel (d) shows that $v_e$ approaches $v_0$ with $\sqrt{N}$ in PRW region. Panel (e) shows that effective persistence time increases increase with $N$ in RUN region. This last quantity is responsible for higher diffusion constant observed since $v_e$ is very close to $v_0$ for all $N$, as seen in panel (a). Panel (f) summarize the combined effect of both parameters, $\tau_e$ and $v_e$, in diffusion constant as function of $\tau$ and $N$.
}
\label{fig.size}
\end{figure}

We address the relationship between the ring's size, $N$, and the diffusion coefficient for the different motion states.
Figures~\ref{fig.size}a(RUN),~\ref{fig.size}b (threshold) and~\ref{fig.size}c (PRW) show the $MSD$ for different ring sizes (N=20,50,100) using parameter $\tau$ to control the system's motion state at constant $Pe$. We exclude the ROT state in this analysis. At low $\tau$ values we find the RUN state, the transition to collective motion is set at $\tau\sim 5$, and above it we reach the PRW state.
All curves present the form described by  Eq.~\ref{msd}, allowing us to fit the effective parameters.

When in RUN state (Fig.~\ref{fig.size}a), all curves in the ballistic interval collapse, indicating the same effective self-propelling speed, $v_e$, with value close to  $v_0$ (Fig.~\ref{MSD_Pe4_N20}c). We also observe in the inset of Fig.~\ref{fig.size}a the effective persistence time increasing as $N$ increases. This is also clear in Fig.~\ref{fig.size}e where $\tau_e$ is divided by $N$, and the curves for the different system sizes coincide at small $\tau$ values. The overall result (Eq. \ref{diff_constant_effect}), in this case, is that the diffusion coefficient scales with $N$, as can be checked in Fig.~\ref{fig.size}f at low $Pe$ values.

The $MSD$ for systems in the threshold to collective motion, Fig.~\ref{fig.size}b, show curves  slightly shifted downward as $N$ increases, indicating a correspondent decrease in $v_e$.  On the other side, in this region, $\tau_e$ starts to grow with $N$( Fig.~\ref{fig.size}e), resulting that $D_e$ remains independent of $N$ on the transition, as indicated by the vertical bar in Fig. \ref{fig.size}f.
Beatrici \cite{beatrici2017} previously found this result in the context of cell segregation.

When in PRW state, Fig.~\ref{fig.size}c, systems present the expected behavior: for increasing $N$ values, the CM moves less since particles are uncorrelated. In Fig.~\ref{fig.size}d we multiply the effective velocity squared by $N$ to show that curves for different sizes collapse at high-$\tau$ values. Since $\tau_e$ is close to its single particle value (Table \ref{tab:table1}), following Eq. \ref{diff_constant_effect}, we expect the ring diffusion to scale with the inverse of $N$. This is observed in Fig.~\ref{fig.size}f at large $\tau$ values.

Fig.~\ref{fig.size}f summarizes the dependency of the diffusion constant with the ring size in the different motion states. There is an evident change in behavior with $N$: from a decrease in RW state to an increase in RUN state, crossing a region without dependence (marked with a vertical line) at the onset of collective motion.

\subsection{Active Ring Morphology}
\label{sec.morfo}

 To characterize morphological changes in the shape of the active ring, we use the gyration tensor, $\Re(t)$~\cite{paoluzzi2016shape,tian2017anomalous,wang2019shape}, defined as 
\begin{eqnarray}
\label{tensor_gir}
\Re(t) &=& \frac{1}{N}\sum_{i}^{N} \vec{r}_{i,CM}(t) \otimes \vec{r}_{i,CM}(t),
\end{eqnarray}
where $\vec{r}_{i,CM} = \vec{r}_{i}(t) - \vec{R}_{CM}(t)$ and $\otimes$ is the tensor product. In matrix form,
\begin{eqnarray}
\label{comp_gir}
   \Re(t) &=& \begin{bmatrix}
R_{xx}(t) & R_{xy}(t) \\ 
R_{yx}(t) & R_{yy}(t) 
\end{bmatrix} \, .
\end{eqnarray}
To quantify the ring extension at time $t$, we measure the gyration's squared radius~\cite{paoluzzi2016shape,tian2017anomalous}
\begin{eqnarray}
\label{raio_gir}
R_{g}^{2}(t) \equiv  Tr(\Re(t)) = \lambda_{1}(t) + \lambda_{2}(t),
\end{eqnarray}
where $\lambda_{1}$ and $\lambda_{2}$ are the gyration tensor eigenvalues.
 Another shape  measure is the asphericity~\cite{paoluzzi2016shape,tian2017anomalous}, defined as
\begin{equation}
\label{asphericity}
   A(t) = \frac{(\lambda_{1}(t) - \lambda_{2}(t))^{2}}{(\lambda_{1}(t) + \lambda_{2}(t))^{2}}\;.
\end{equation}
The limiting cases where $A(t) = 0$ and $A(t) = 1$ corresponds to a circle and  to a rod, respectively.
We use the stationary average of those quantities, $\left \langle R_{g} \right \rangle$, $\left \langle A \right \rangle$ and the asphericity fluctuation, $\delta A$, to characterize the ring format. 

\begin{figure}[!h]
\centering
\includegraphics[width=1.0\columnwidth]{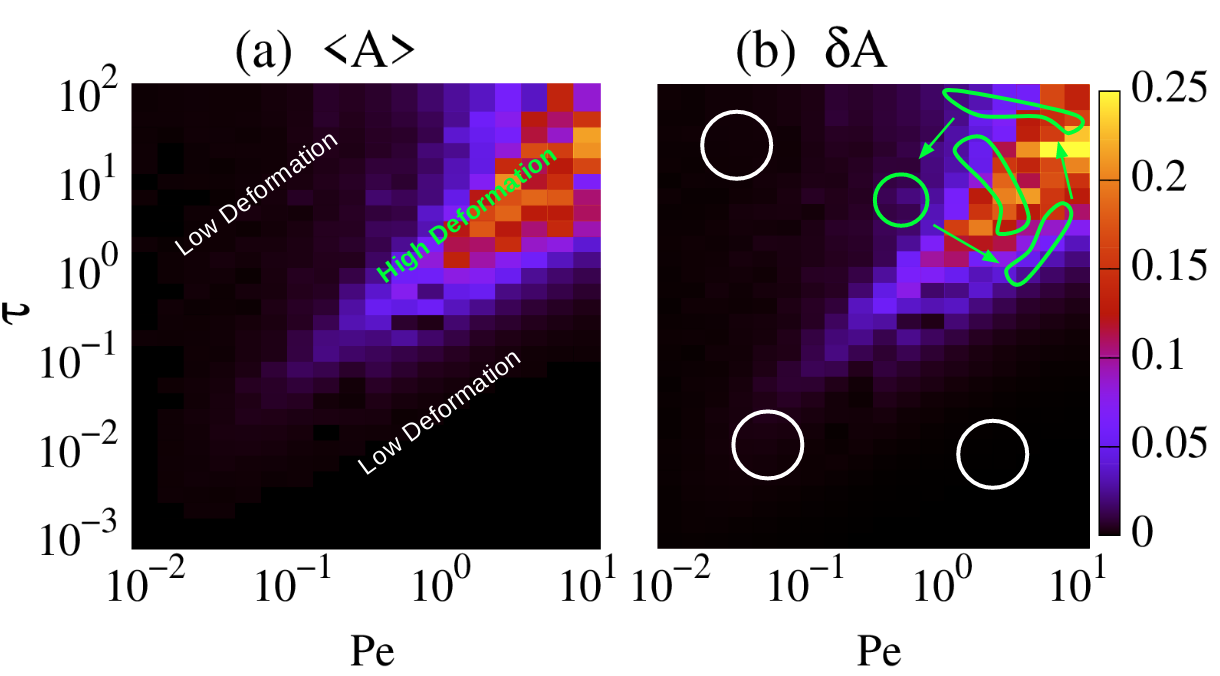} 
\caption{
State diagram  ($Pe \times \tau$) for (a) asphericity, $\left \langle A \right \rangle$, and (b) its fluctuations, $\delta A$, while parameters $B = 0$, $Fn = 1$, $\beta = 0$ and $N=100$ are kept fixed. In RUN state, the shape changes are small, since all particles move aligned with CM, and the ring preserves its initial circular format. In RW state, the particles behave as uncorrelated ones and the ring maintains its initial circular format but, now, with higher membrane fluctuations than those of RUN state.
Note that parameter axes are in log scale but the color bar is in linear scale.
}
\label{phase_structure_N100}
\end{figure}

Fig.~\ref{phase_structure_N100} shows the phase diagram for asphericity and its fluctuation. As in our previous diagram, Fig.~\ref{phasediagram_collectivity_B0_beta0}, we fix the flexure number ($F_n=1$) and vary parameters $\tau$ and $Pe$. On top of that, we sketch illustrative ring formats. We choose $N=100$ because the previously used ring's size, $N=20$, shows small shape variations for this flexure number. 

At high $\tau$ and low $Pe$ values (Fig.~\ref{phase_structure_N100}), the ring maintains an almost circular format, that is, $A(t) \rightarrow 0$ and $\delta A \rightarrow 0$. In this limit, the active particles in the ring show short characteristic persistence time, and each particle quickly changes its direction, resulting in a mean circular shape with the boundary fluctuating at small scales.

The circular shape is also stable in the collective motion region ($\tau/Pe\ll1$). Here, the alignment is responsible for moving particles in the same orientation with the same self-propelled speed. 
The boundary fluctuates less than in the previous case.

As both $\tau$ and $Pe$ increase, the ring shape changes from an almost circular format to an elongated one. This change implies an increase in both $\left \langle A \right \rangle$ and $\delta A$, as shown along the main diagonal of 
Fig.~\ref{phase_structure_N100}.
In this region, the active particles do not align globally with each other (absence of collective motion). Still, similar to what happens in the Run and Rotate state, the observed characteristic persistence time is high enough to ensure subgroups moving persistently in different directions, causing ring deformation.
The eigenvectors' direction of $\Re$ fluctuate with time, and while changing direction, the system spends some time close to a  circular format. These low contributions of the circular shape to $\left \langle A \right \rangle$ are the reason for the low values observed in Fig.~\ref{phase_structure_N100}  despite the elongated aspect. 
These fluctuations are responsible  for higher values for $\delta A$  measured in our simulations, see Fig.~\ref{phase_structure_N100}b.

\begin{figure}[!h]
\centering
\includegraphics[width=9.27cm]{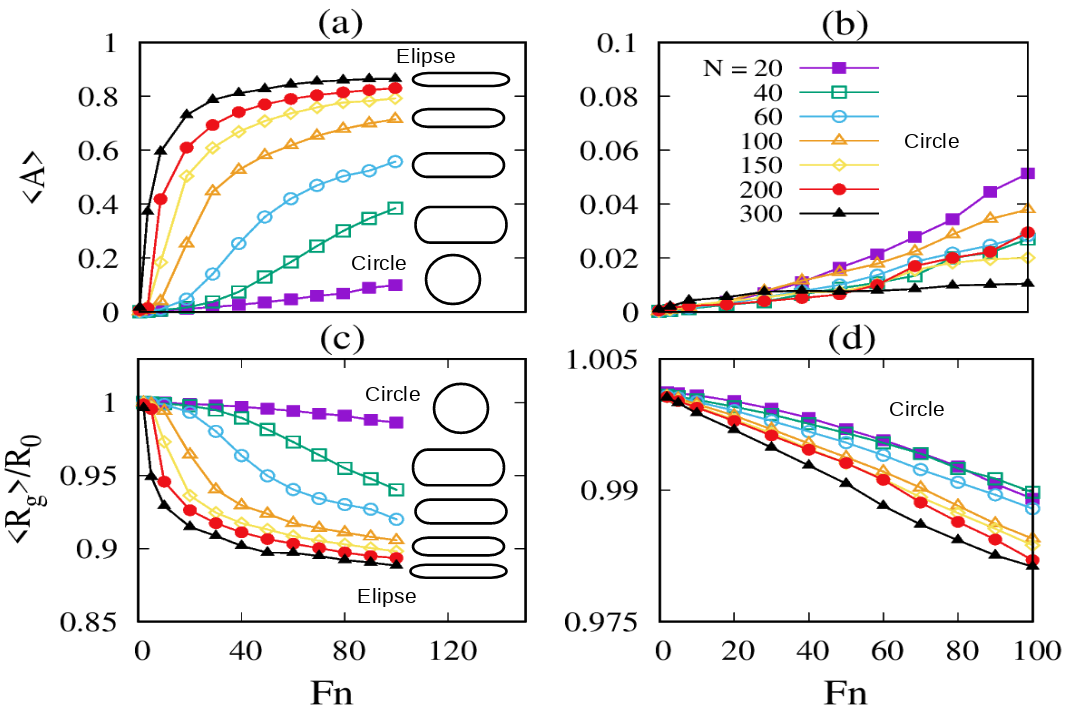} 
\caption{
Observed changes in the shape of the ring in collective motion for RUN state, left panels, and ROT state, right panels, for set of parameters: $Pe = 5$, $B = 0$ and $\tau = 0.1$. 
Left panels show simulation results for (a) average asphericity and (c) ratio between the average radius of gyration and initial radius of gyration $R_{0}$ as a function of flexure number for different sizes of the ring, $N$, and initialization with $\beta = 0$. Both figures indicate that larger systems present a more pronounced modification of their shape as systems become more flexible. It goes from a circular format when $N=20$ to an elongated one, resembling an ellipse, for $N=300$. 
Right panels (b) and (d) show same quantities for an initialization with $\beta = 1$. When the system is in ROT state there is only a small departure of the initial circular format independent of the ring's flexibility and size.
}
\label{asphericity_radius}
\end{figure}

\begin{figure}[!h]
\centering
\includegraphics[width=8.27cm]{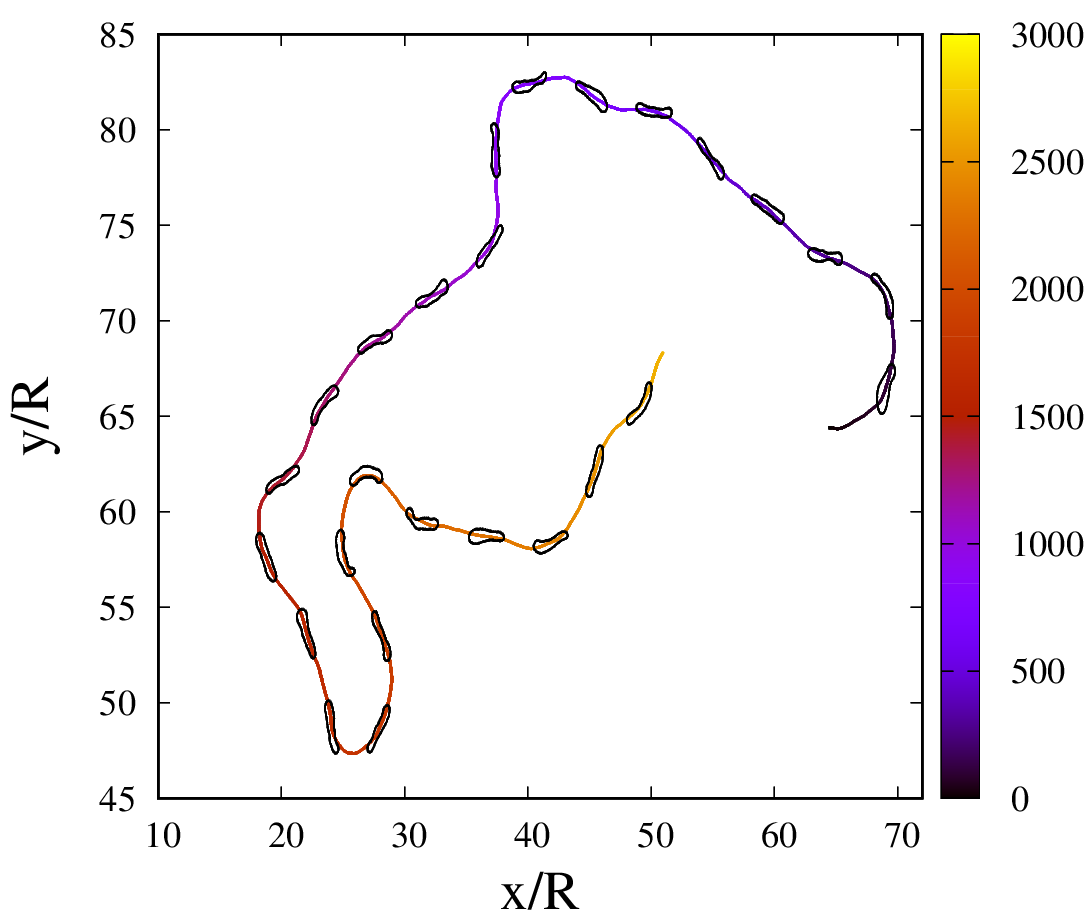}
\caption{
Center of Mass trajectory of the active ring for a simulation sample in RUN motion state with parameters: $N = 100$, $Pe = 5$, $Fn = 100$, $\beta = 0$, $\tau = 0.1$ and $B = 0$. On top of the trajectory of CM, we show snapshots of the active ring where it is possible to note that: (i) ring has an elongated format; and (ii) it always moves in the most elongated direction. These observations suggest the presence of spontaneous global polarization of the active ring. Colors indicate time (see color bar). See Movie5 in Supl. Material.
}
\label{trajectory}
\end{figure}

To address the variation of shape with rigidity, we focus on the region of collective motion. 
As the bending constant, responsible for ring's  rigidity, decreases (flexure number increases), the ring remains in an elongated format. Measuring the asphericity (Fig.~\ref{asphericity_radius}a) we find that the flattening is more evident as the flexure number and the number of particles increase, thus, $\left \langle A \right\rangle\rightarrow 1$. In Fig.~\ref{asphericity_radius}c we compare the gyration radio $\left \langle R_{g} \right \rangle$ with the one from a circle, $R_0$, for different flexure number and number of particles. For large values of $Fn$ and $N$ we observe an extremely deformed active ring (black curve in the Fig.~\ref{asphericity_radius}c), in that case we find $\left \langle R_{g} \right \rangle/R_{0} \sim 0.9$. This value is close to the one found for a rod,
$\left \langle R_{g} \right \rangle/R_{0}$, which converges to  $\pi/2\sqrt{3} = 0.90689$ as $N$ increases (Supp. Mat.). 

When the system is in the ROT state, with an initial circular condition and $\beta=1$, particle velocities are tangential to the ring, causing no observable changes in its shape even when increasing the flexure number (Figs.~\ref{asphericity_radius}b and d).

Finally, when the system is in RUN state and presents high deformation, the ring becomes flattened. In that case, the collective movement has a well-defined direction oriented parallel to the larger ring dimension, which we may interpret as a spontaneous emergence of a polarization direction. Whenever there is a change in the collective movement, there is a realignment of that largest dimension. This behavior is illustrated in Fig.~\ref{trajectory}, where ring snapshots are plotted along its trajectory. Colors indicate the time. The causes for this emerging polarization remain an open question.

\section{Summary and conclusions}
\label{sec.conc}

In this work, we study a model for a ring composed of active particles. 
We establish the conditions for the emergence of collective movement and study the ring's deformation by tuning system parameters such as rigidity constant, angular and translational noises, and angular relaxation time.
We can identify  different ring motion states. In the limit $\tau/Pe\gg1$, we observe a behavior compatible with an ABP system, which results in an $MSD$ for the center of mass equivalent to a system of $N$ interconnected particles subjected to random independent noise. When $\tau/Pe\ll 1$, the correlation between particles becomes pronounced, resulting in two forms of collective movement: collective translational movement (RUN), where particles move with their speeds nearly parallel; and collective rotational movement (ROT), where particles rotate around the CM. As far as we know, such rotational state has never been observed in single cell experiments, and is probably due to an excess in degrees of freedom if compared, for example, with the possible orientations found in actomyosin fibers\cite{Gunning2009}. For small rings, $N < 200$, we identify a dynamics where the system alternates between these states of collective movement (RRM). As the ring's size increases, the transitions between collective states of motion become unlikely.

All simulations show $MSD$ measurements compatible with the ABP limit known solution. That is, ballistic for short times, followed by diffusion. The ballistic regime is preceded by a diffusive one only when in the presence of translational noise~\cite{mandal2020,THOMAS2020124493}. Even at high flexure numbers, when membrane oscillations are large, angular noise by itself cannot produce a short time diffusive behavior. Also, we found no super-diffusive regime intermediary between the ballistic and the diffusive~\cite{velasco2017complex,Potdar2009} ones. 

We fit effective self-propulsion speed, $v_e$, and persistence time, $\tau_e$, based on the analytic results' functional form.  The ROT state's fitting procedure resulted in the lowest values for the $\tau_e$, and $v_e$ since particles circulate the CM without generating a significant displacement. On large time scales, we observe that the CM performs a diffusive process. The RUN state presented the highest values for the effective parameters, with particles moving aligned with each other in a movement with considerable temporal persistence and, consequently, large $MSD$. The RRM state's fit presents intermediary parameters since they depend on the fraction of time spent in the RUN state, with the ROT state poorly contributing to CM displacement. Another interesting remark is that even a small translational noise undermines the onset to collective movement deviating the transition to an order of magnitude lower $\tau$ values.

By varying the number of particles, $N$,  we note a shift in the effective parameters fitted for the $MSD$. As $N$ increases in the RUN state, the effective self-propulsion speed, $v_e$, approaches $v_0$, and the diffusion coefficient increases linearly with it. When in the threshold between the collective movement and the PRW phase, fitted parameters are $N$ independent. In the PRW state, both parameters decrease with $N$ as expected for particles dominated by uncorrelated noise.  A theoretical approach would help shed light on $N$'s effective parameters dependence at the transition to the collective movement and above it. A previous mean cluster study~\cite{beatrici2017} relating Vicsek's collective movement parameter~\cite{vicsek} and diffusion coefficient mass dependence correctly explained cell segregation time scales. However, it offered no theoretical hint for that relation.

In the ring morphology study, we found that the ring maintains its initial circular shape when $\tau$ or $Pe$ is low enough. It presents large fluctuations in shape when $\tau$ and $Pe$ increase. Larger rings present a substantial variation in the possible shapes they can assume when decreasing the curvature potential's stiffness. 
Large soft rings assume a slug-like form when in collective motion, with a well-defined movement polarization along the largest ring direction. Since we deal with correlated active particles, the emergence of polarization, in this case, can be interpreted as a nonlinear instability of the center of mass, as proposed by Blanch-Mercader and coworkers\cite{Blan2013}, but the relation between the slug larger direction and the global velocity remains an open question.

The active ring system proposed aims to serve as a model for cells. Many models in Active Matter are single particle-based and unable to describe cells' morphological properties. Furthermore, for being a bead-spring model, its use in phenomena such as durotaxis, chemotaxis, cell segregation, cell crawling, or wound healing is easy to implement with modest modifications by including interaction forces among cells or external chemical fields. Here, in this first work, we characterize the dynamic and morphological properties of a single ring. In future works, we will study systems composed of many of these.

\section*{Conflicts of interest}
There are no conflicts to declare

\section*{Acknowledgements}
This work is dedicated to the memory of Cássio Kirch. 
E.F.T. thanks the Brazilian funding agencies CNPq and Capes. H.C.M.F. acknowledges  Universitat de Barcelona where  part  of  this  work was developed. L.G.B. acknowledges the Max-Planck Institute of Ploen, where part of this work was developed. All authors acknowledge the suggestions and discussions with S. Lira. The simulations were performed on the IF-UFRGS computing cluster infrastructure.  



\balance



\begin{mcitethebibliography}{44}
\providecommand*{\natexlab}[1]{#1}
\providecommand*{\mciteSetBstSublistMode}[1]{}
\providecommand*{\mciteSetBstMaxWidthForm}[2]{}
\providecommand*{\mciteBstWouldAddEndPuncttrue}
  {\def\EndOfBibitem{\unskip.}}
\providecommand*{\mciteBstWouldAddEndPunctfalse}
  {\let\EndOfBibitem\relax}
\providecommand*{\mciteSetBstMidEndSepPunct}[3]{}
\providecommand*{\mciteSetBstSublistLabelBeginEnd}[3]{}
\providecommand*{\EndOfBibitem}{}
\mciteSetBstSublistMode{f}
\mciteSetBstMaxWidthForm{subitem}
{(\emph{\alph{mcitesubitemcount}})}
\mciteSetBstSublistLabelBeginEnd{\mcitemaxwidthsubitemform\space}
{\relax}{\relax}

\bibitem[Ramaswamy(2010)]{rama2010}
S.~Ramaswamy, \emph{Annual Review of Condensed Matter Physics}, 2010,
  \textbf{1}, 323--345\relax
\mciteBstWouldAddEndPuncttrue
\mciteSetBstMidEndSepPunct{\mcitedefaultmidpunct}
{\mcitedefaultendpunct}{\mcitedefaultseppunct}\relax
\EndOfBibitem
\bibitem[Levin(2012)]{LEVIN2012243}
M.~Levin, \emph{Biosystems}, 2012, \textbf{109}, 243 -- 261\relax
\mciteBstWouldAddEndPuncttrue
\mciteSetBstMidEndSepPunct{\mcitedefaultmidpunct}
{\mcitedefaultendpunct}{\mcitedefaultseppunct}\relax
\EndOfBibitem
\bibitem[Cochet-Escartin \emph{et~al.}(2017)Cochet-Escartin, Locke, Shi,
  Steele, and Collins]{COCHETESCARTIN20172827}
O.~Cochet-Escartin, T.~T. Locke, W.~H. Shi, R.~E. Steele and E.-M.~S. Collins,
  \emph{Biophysical Journal}, 2017, \textbf{113}, 2827 -- 2841\relax
\mciteBstWouldAddEndPuncttrue
\mciteSetBstMidEndSepPunct{\mcitedefaultmidpunct}
{\mcitedefaultendpunct}{\mcitedefaultseppunct}\relax
\EndOfBibitem
\bibitem[Marchetti \emph{et~al.}(2013)Marchetti, Joanny, Ramaswamy, Liverpool,
  Prost, Rao, and Simha]{marcheti2013}
M.~C. Marchetti, J.~F. Joanny, S.~Ramaswamy, T.~B. Liverpool, J.~Prost, M.~Rao
  and R.~A. Simha, \emph{Rev. Mod. Phys.}, 2013, \textbf{85}, 1143--1189\relax
\mciteBstWouldAddEndPuncttrue
\mciteSetBstMidEndSepPunct{\mcitedefaultmidpunct}
{\mcitedefaultendpunct}{\mcitedefaultseppunct}\relax
\EndOfBibitem
\bibitem[Van~Liedekerke \emph{et~al.}(2015)Van~Liedekerke, Palm, Jagiella, and
  Drasdo]{vanliedekerke2015}
P.~Van~Liedekerke, M.~M. Palm, N.~Jagiella and D.~Drasdo, \emph{Computational
  Particle Mechanics}, 2015, \textbf{2}, 401--444\relax
\mciteBstWouldAddEndPuncttrue
\mciteSetBstMidEndSepPunct{\mcitedefaultmidpunct}
{\mcitedefaultendpunct}{\mcitedefaultseppunct}\relax
\EndOfBibitem
\bibitem[Steinberg(1963)]{Steinberg401}
M.~S. Steinberg, \emph{Science}, 1963, \textbf{141}, 401--408\relax
\mciteBstWouldAddEndPuncttrue
\mciteSetBstMidEndSepPunct{\mcitedefaultmidpunct}
{\mcitedefaultendpunct}{\mcitedefaultseppunct}\relax
\EndOfBibitem
\bibitem[Jones \emph{et~al.}(1989)Jones, Evans, and Lee]{Jones1989}
B.~M. Jones, P.~M. Evans and D.~A. Lee, \emph{Experimental Cell Research},
  1989, \textbf{180}, 287 -- 296\relax
\mciteBstWouldAddEndPuncttrue
\mciteSetBstMidEndSepPunct{\mcitedefaultmidpunct}
{\mcitedefaultendpunct}{\mcitedefaultseppunct}\relax
\EndOfBibitem
\bibitem[Belmonte \emph{et~al.}(2008)Belmonte, Thomas, Brunnet, de~Almeida, and
  Chat\'e]{belmonte2008}
J.~M. Belmonte, G.~L. Thomas, L.~G. Brunnet, R.~M.~C. de~Almeida and
  H.~Chat\'e, \emph{Phys. Rev. Lett.}, 2008, \textbf{100}, 248702\relax
\mciteBstWouldAddEndPuncttrue
\mciteSetBstMidEndSepPunct{\mcitedefaultmidpunct}
{\mcitedefaultendpunct}{\mcitedefaultseppunct}\relax
\EndOfBibitem
\bibitem[Beatrici and Brunnet(2011)]{beatrici2011}
C.~P. Beatrici and L.~G. Brunnet, \emph{Phys. Rev. E}, 2011, \textbf{84},
  031927\relax
\mciteBstWouldAddEndPuncttrue
\mciteSetBstMidEndSepPunct{\mcitedefaultmidpunct}
{\mcitedefaultendpunct}{\mcitedefaultseppunct}\relax
\EndOfBibitem
\bibitem[Harris(1976)]{harris1976}
A.~K. Harris, \emph{Journal of Theoretical Biology}, 1976, \textbf{61}, 267 --
  285\relax
\mciteBstWouldAddEndPuncttrue
\mciteSetBstMidEndSepPunct{\mcitedefaultmidpunct}
{\mcitedefaultendpunct}{\mcitedefaultseppunct}\relax
\EndOfBibitem
\bibitem[Abercrombie(1980)]{Aber1980}
M.~Abercrombie, \emph{Proceedings of the Royal Society of London. Series B.
  Biological Sciences}, 1980, \textbf{207}, 129--147\relax
\mciteBstWouldAddEndPuncttrue
\mciteSetBstMidEndSepPunct{\mcitedefaultmidpunct}
{\mcitedefaultendpunct}{\mcitedefaultseppunct}\relax
\EndOfBibitem
\bibitem[Schwarz and Safran(2013)]{Schwarz2013}
U.~S. Schwarz and S.~A. Safran, \emph{Rev. Mod. Phys.}, 2013, \textbf{85},
  1327--1381\relax
\mciteBstWouldAddEndPuncttrue
\mciteSetBstMidEndSepPunct{\mcitedefaultmidpunct}
{\mcitedefaultendpunct}{\mcitedefaultseppunct}\relax
\EndOfBibitem
\bibitem[Graner and Glazier(1992)]{graner1992}
F.~m.~c. Graner and J.~A. Glazier, \emph{Phys. Rev. Lett.}, 1992, \textbf{69},
  2013--2016\relax
\mciteBstWouldAddEndPuncttrue
\mciteSetBstMidEndSepPunct{\mcitedefaultmidpunct}
{\mcitedefaultendpunct}{\mcitedefaultseppunct}\relax
\EndOfBibitem
\bibitem[Bi \emph{et~al.}(2016)Bi, Yang, Marchetti, and Manning]{bi2016}
D.~Bi, X.~Yang, M.~C. Marchetti and M.~L. Manning, \emph{Phys. Rev. X}, 2016,
  \textbf{6}, 021011\relax
\mciteBstWouldAddEndPuncttrue
\mciteSetBstMidEndSepPunct{\mcitedefaultmidpunct}
{\mcitedefaultendpunct}{\mcitedefaultseppunct}\relax
\EndOfBibitem
\bibitem[Barton \emph{et~al.}(2017)Barton, Henkes, Weijer, and
  Sknepnek]{barton2017}
D.~L. Barton, S.~Henkes, C.~J. Weijer and R.~Sknepnek, \emph{PLOS Computational
  Biology}, 2017, \textbf{13}, 1--34\relax
\mciteBstWouldAddEndPuncttrue
\mciteSetBstMidEndSepPunct{\mcitedefaultmidpunct}
{\mcitedefaultendpunct}{\mcitedefaultseppunct}\relax
\EndOfBibitem
\bibitem[Shao \emph{et~al.}(2012)Shao, Levine, and Rappel]{shao2012}
D.~Shao, H.~Levine and W.-J. Rappel, \emph{Proceedings of the National Academy
  of Sciences}, 2012, \textbf{109}, 6851--6856\relax
\mciteBstWouldAddEndPuncttrue
\mciteSetBstMidEndSepPunct{\mcitedefaultmidpunct}
{\mcitedefaultendpunct}{\mcitedefaultseppunct}\relax
\EndOfBibitem
\bibitem[Blanch-Mercader and Casademunt(2013)]{Blan2013}
C.~Blanch-Mercader and J.~Casademunt, \emph{Phys. Rev. Lett.}, 2013,
  \textbf{110}, 078102\relax
\mciteBstWouldAddEndPuncttrue
\mciteSetBstMidEndSepPunct{\mcitedefaultmidpunct}
{\mcitedefaultendpunct}{\mcitedefaultseppunct}\relax
\EndOfBibitem
\bibitem[Vicsek \emph{et~al.}(1995)Vicsek, Czir\'ok, Ben-Jacob, Cohen, and
  Shochet]{vicsek}
T.~Vicsek, A.~Czir\'ok, E.~Ben-Jacob, I.~Cohen and O.~Shochet, \emph{Phys. Rev.
  Lett.}, 1995, \textbf{75}, 1226--1229\relax
\mciteBstWouldAddEndPuncttrue
\mciteSetBstMidEndSepPunct{\mcitedefaultmidpunct}
{\mcitedefaultendpunct}{\mcitedefaultseppunct}\relax
\EndOfBibitem
\bibitem[Erdmann \emph{et~al.}(2005)Erdmann, Ebeling, and
  Mikhailov]{erdmann2005noise}
U.~Erdmann, W.~Ebeling and A.~S. Mikhailov, \emph{Physical Review E}, 2005,
  \textbf{71}, 051904\relax
\mciteBstWouldAddEndPuncttrue
\mciteSetBstMidEndSepPunct{\mcitedefaultmidpunct}
{\mcitedefaultendpunct}{\mcitedefaultseppunct}\relax
\EndOfBibitem
\bibitem[Cates and Tailleur(2013)]{cates2013}
M.~E. Cates and J.~Tailleur, \emph{{EPL} (Europhysics Letters)}, 2013,
  \textbf{101}, 20010\relax
\mciteBstWouldAddEndPuncttrue
\mciteSetBstMidEndSepPunct{\mcitedefaultmidpunct}
{\mcitedefaultendpunct}{\mcitedefaultseppunct}\relax
\EndOfBibitem
\bibitem[Howse \emph{et~al.}(2007)Howse, Jones, Ryan, Gough, Vafabakhsh, and
  Golestanian]{howse2007}
J.~R. Howse, R.~A.~L. Jones, A.~J. Ryan, T.~Gough, R.~Vafabakhsh and
  R.~Golestanian, \emph{Phys. Rev. Lett.}, 2007, \textbf{99}, 048102\relax
\mciteBstWouldAddEndPuncttrue
\mciteSetBstMidEndSepPunct{\mcitedefaultmidpunct}
{\mcitedefaultendpunct}{\mcitedefaultseppunct}\relax
\EndOfBibitem
\bibitem[Purcell(1977)]{purcell1977life}
E.~M. Purcell, \emph{American journal of physics}, 1977, \textbf{45},
  3--11\relax
\mciteBstWouldAddEndPuncttrue
\mciteSetBstMidEndSepPunct{\mcitedefaultmidpunct}
{\mcitedefaultendpunct}{\mcitedefaultseppunct}\relax
\EndOfBibitem
\bibitem[Bechinger \emph{et~al.}(2016)Bechinger, Di~Leonardo, L{\"o}wen,
  Reichhardt, Volpe, and Volpe]{bechinger2016active}
C.~Bechinger, R.~Di~Leonardo, H.~L{\"o}wen, C.~Reichhardt, G.~Volpe and
  G.~Volpe, \emph{Reviews of Modern Physics}, 2016, \textbf{88}, 045006\relax
\mciteBstWouldAddEndPuncttrue
\mciteSetBstMidEndSepPunct{\mcitedefaultmidpunct}
{\mcitedefaultendpunct}{\mcitedefaultseppunct}\relax
\EndOfBibitem
\bibitem[Szab\'o \emph{et~al.}(2006)Szab\'o, Sz\"oll\"osi, G\"onci, Jur\'anyi,
  Selmeczi, and Vicsek]{szabo}
B.~Szab\'o, G.~J. Sz\"oll\"osi, B.~G\"onci, Z.~Jur\'anyi, D.~Selmeczi and
  T.~Vicsek, \emph{Phys. Rev. E}, 2006, \textbf{74}, 061908\relax
\mciteBstWouldAddEndPuncttrue
\mciteSetBstMidEndSepPunct{\mcitedefaultmidpunct}
{\mcitedefaultendpunct}{\mcitedefaultseppunct}\relax
\EndOfBibitem
\bibitem[Allen and Tildesley(2017)]{allen2017}
M.~Allen and D.~Tildesley, \emph{Computer Simulation of Liquids}, Oxford
  University Press, 2017\relax
\mciteBstWouldAddEndPuncttrue
\mciteSetBstMidEndSepPunct{\mcitedefaultmidpunct}
{\mcitedefaultendpunct}{\mcitedefaultseppunct}\relax
\EndOfBibitem
\bibitem[Mart{\'\i}n-G{\'o}mez \emph{et~al.}(2018)Mart{\'\i}n-G{\'o}mez, Levis,
  D{\'\i}az-Guilera, and Pagonabarraga]{martin2018collective}
A.~Mart{\'\i}n-G{\'o}mez, D.~Levis, A.~D{\'\i}az-Guilera and I.~Pagonabarraga,
  \emph{Soft matter}, 2018, \textbf{14}, 2610--2618\relax
\mciteBstWouldAddEndPuncttrue
\mciteSetBstMidEndSepPunct{\mcitedefaultmidpunct}
{\mcitedefaultendpunct}{\mcitedefaultseppunct}\relax
\EndOfBibitem
\bibitem[Duman \emph{et~al.}(2018)Duman, Isele-Holder, Elgeti, and
  Gompper]{duman2018collective}
{\"O}.~Duman, R.~E. Isele-Holder, J.~Elgeti and G.~Gompper, \emph{Soft Matter},
  2018, \textbf{14}, 4483--4494\relax
\mciteBstWouldAddEndPuncttrue
\mciteSetBstMidEndSepPunct{\mcitedefaultmidpunct}
{\mcitedefaultendpunct}{\mcitedefaultseppunct}\relax
\EndOfBibitem
\bibitem[Czir{\'o}k \emph{et~al.}(1996)Czir{\'o}k, Ben-Jacob, Cohen, and
  Vicsek]{czirok1996formation}
A.~Czir{\'o}k, E.~Ben-Jacob, I.~Cohen and T.~Vicsek, \emph{Physical Review E},
  1996, \textbf{54}, 1791\relax
\mciteBstWouldAddEndPuncttrue
\mciteSetBstMidEndSepPunct{\mcitedefaultmidpunct}
{\mcitedefaultendpunct}{\mcitedefaultseppunct}\relax
\EndOfBibitem
\bibitem[Str{\"o}mbom(2011)]{strombom2011collective}
D.~Str{\"o}mbom, \emph{Journal of theoretical biology}, 2011, \textbf{283},
  145--151\relax
\mciteBstWouldAddEndPuncttrue
\mciteSetBstMidEndSepPunct{\mcitedefaultmidpunct}
{\mcitedefaultendpunct}{\mcitedefaultseppunct}\relax
\EndOfBibitem
\bibitem[Schweitzer(2003)]{schweitzer2003brownian}
F.~Schweitzer, \emph{Brownian agents and active particles: collective dynamics
  in the natural and social sciences}, Springer Science \& Business Media,
  2003\relax
\mciteBstWouldAddEndPuncttrue
\mciteSetBstMidEndSepPunct{\mcitedefaultmidpunct}
{\mcitedefaultendpunct}{\mcitedefaultseppunct}\relax
\EndOfBibitem
\bibitem[Romanczuk \emph{et~al.}(2012)Romanczuk, B{\"a}r, Ebeling, Lindner, and
  Schimansky-Geier]{romanczuk2012active}
P.~Romanczuk, M.~B{\"a}r, W.~Ebeling, B.~Lindner and L.~Schimansky-Geier,
  \emph{The European Physical Journal Special Topics}, 2012, \textbf{202},
  1--162\relax
\mciteBstWouldAddEndPuncttrue
\mciteSetBstMidEndSepPunct{\mcitedefaultmidpunct}
{\mcitedefaultendpunct}{\mcitedefaultseppunct}\relax
\EndOfBibitem
\bibitem[Digregorio \emph{et~al.}(2018)Digregorio, Levis, Suma, Cugliandolo,
  Gonnella, and Pagonabarraga]{digregorio2018}
P.~Digregorio, D.~Levis, A.~Suma, L.~F. Cugliandolo, G.~Gonnella and
  I.~Pagonabarraga, \emph{Phys. Rev. Lett.}, 2018, \textbf{121}, 098003\relax
\mciteBstWouldAddEndPuncttrue
\mciteSetBstMidEndSepPunct{\mcitedefaultmidpunct}
{\mcitedefaultendpunct}{\mcitedefaultseppunct}\relax
\EndOfBibitem
\bibitem[Gal \emph{et~al.}(2013)Gal, Lechtman-Goldstein, and Weihs]{Gal2013}
N.~Gal, D.~Lechtman-Goldstein and D.~Weihs, \emph{Rheologica Acta}, 2013,
  \textbf{52}, 425--443\relax
\mciteBstWouldAddEndPuncttrue
\mciteSetBstMidEndSepPunct{\mcitedefaultmidpunct}
{\mcitedefaultendpunct}{\mcitedefaultseppunct}\relax
\EndOfBibitem
\bibitem[Basu \emph{et~al.}(2018)Basu, Majumdar, Rosso, and
  Schehr]{basu2018active}
U.~Basu, S.~N. Majumdar, A.~Rosso and G.~Schehr, \emph{Physical Review E},
  2018, \textbf{98}, 062121\relax
\mciteBstWouldAddEndPuncttrue
\mciteSetBstMidEndSepPunct{\mcitedefaultmidpunct}
{\mcitedefaultendpunct}{\mcitedefaultseppunct}\relax
\EndOfBibitem
\bibitem[Fodor and Marchetti(2018)]{fodor2018statistical}
{\'E}.~Fodor and M.~C. Marchetti, \emph{Physica A: Statistical Mechanics and
  its Applications}, 2018, \textbf{504}, 106--120\relax
\mciteBstWouldAddEndPuncttrue
\mciteSetBstMidEndSepPunct{\mcitedefaultmidpunct}
{\mcitedefaultendpunct}{\mcitedefaultseppunct}\relax
\EndOfBibitem
\bibitem[Beatrici \emph{et~al.}(2017)Beatrici, de~Almeida, and
  Brunnet]{beatrici2017}
C.~P. Beatrici, R.~M.~C. de~Almeida and L.~G. Brunnet, \emph{Phys. Rev. E},
  2017, \textbf{95}, 032402\relax
\mciteBstWouldAddEndPuncttrue
\mciteSetBstMidEndSepPunct{\mcitedefaultmidpunct}
{\mcitedefaultendpunct}{\mcitedefaultseppunct}\relax
\EndOfBibitem
\bibitem[Paoluzzi \emph{et~al.}(2016)Paoluzzi, Di~Leonardo, Marchetti, and
  Angelani]{paoluzzi2016shape}
M.~Paoluzzi, R.~Di~Leonardo, M.~C. Marchetti and L.~Angelani, \emph{Scientific
  reports}, 2016, \textbf{6}, 34146\relax
\mciteBstWouldAddEndPuncttrue
\mciteSetBstMidEndSepPunct{\mcitedefaultmidpunct}
{\mcitedefaultendpunct}{\mcitedefaultseppunct}\relax
\EndOfBibitem
\bibitem[Tian \emph{et~al.}(2017)Tian, Gu, Guo, and Chen]{tian2017anomalous}
W.-D. Tian, Y.~Gu, Y.-K. Guo and K.~Chen, \emph{Chinese Physics B}, 2017,
  \textbf{26}, 100502\relax
\mciteBstWouldAddEndPuncttrue
\mciteSetBstMidEndSepPunct{\mcitedefaultmidpunct}
{\mcitedefaultendpunct}{\mcitedefaultseppunct}\relax
\EndOfBibitem
\bibitem[Wang \emph{et~al.}(2019)Wang, Guo, Tian, and Chen]{wang2019shape}
C.~Wang, Y.-k. Guo, W.-d. Tian and K.~Chen, \emph{The Journal of chemical
  physics}, 2019, \textbf{150}, 044907\relax
\mciteBstWouldAddEndPuncttrue
\mciteSetBstMidEndSepPunct{\mcitedefaultmidpunct}
{\mcitedefaultendpunct}{\mcitedefaultseppunct}\relax
\EndOfBibitem
\bibitem[Gunning \emph{et~al.}(2015)Gunning, Ghoshdastider, Whitaker, Popp, and
  Robinson]{Gunning2009}
P.~W. Gunning, U.~Ghoshdastider, S.~Whitaker, D.~Popp and R.~C. Robinson,
  \emph{Journal of Cell Science}, 2015, \textbf{128}, 2009--2019\relax
\mciteBstWouldAddEndPuncttrue
\mciteSetBstMidEndSepPunct{\mcitedefaultmidpunct}
{\mcitedefaultendpunct}{\mcitedefaultseppunct}\relax
\EndOfBibitem
\bibitem[Mandal \emph{et~al.}(2020)Mandal, Kurzthaler, Franosch, and
  L\"owen]{mandal2020}
S.~Mandal, C.~Kurzthaler, T.~Franosch and H.~L\"owen, \emph{Phys. Rev. Lett.},
  2020, \textbf{125}, 138002\relax
\mciteBstWouldAddEndPuncttrue
\mciteSetBstMidEndSepPunct{\mcitedefaultmidpunct}
{\mcitedefaultendpunct}{\mcitedefaultseppunct}\relax
\EndOfBibitem
\bibitem[Thomas \emph{et~al.}(2020)Thomas, Fortuna, Perrone, Glazier, Belmonte,
  and de~Alm~eida]{THOMAS2020124493}
G.~L. Thomas, I.~Fortuna, G.~C. Perrone, J.~A. Glazier, J.~M. Belmonte and
  R.~M. de~Alm~eida, \emph{Physica A: Statistical Mechanics and its
  Applications}, 2020, \textbf{550}, 124493\relax
\mciteBstWouldAddEndPuncttrue
\mciteSetBstMidEndSepPunct{\mcitedefaultmidpunct}
{\mcitedefaultendpunct}{\mcitedefaultseppunct}\relax
\EndOfBibitem
\bibitem[Velasco \emph{et~al.}(2017)Velasco, Ghahnaviyeh, Pishkenari, Auth, and
  Gompper]{velasco2017complex}
C.~A. Velasco, S.~D. Ghahnaviyeh, H.~N. Pishkenari, T.~Auth and G.~Gompper,
  \emph{Soft Matter}, 2017, \textbf{13}, 5865--5876\relax
\mciteBstWouldAddEndPuncttrue
\mciteSetBstMidEndSepPunct{\mcitedefaultmidpunct}
{\mcitedefaultendpunct}{\mcitedefaultseppunct}\relax
\EndOfBibitem
\bibitem[Potdar \emph{et~al.}(2009)Potdar, Lu, Jeon, Weaver, and
  Cummings]{Potdar2009}
A.~Potdar, J.~Lu, J.~Jeon, A.~Weaver and P.~Cummings, \emph{Ann Biomed Eng.},
  2009, \textbf{37}, 230\relax
\mciteBstWouldAddEndPuncttrue
\mciteSetBstMidEndSepPunct{\mcitedefaultmidpunct}
{\mcitedefaultendpunct}{\mcitedefaultseppunct}\relax
\EndOfBibitem
\end{mcitethebibliography}
\bibliographystyle{rsc} 

\providecommand*{\mcitethebibliography}{\thebibliography}
\csname @ifundefined\endcsname{endmcitethebibliography}
{\let\endmcitethebibliography\endthebibliography}{}

\end{document}